\documentclass[journal]{IEEEtran}
\usepackage{stfloats}
\usepackage{color}
\usepackage{cite}
\usepackage{epsfig}
\usepackage{float}
\usepackage{subfigure}
\usepackage{times}
\usepackage{amsmath}
\newtheorem{theorem}{Theorem}
\newtheorem{lemma}{Lemma}
\usepackage{amssymb}
\usepackage{multirow}
\usepackage{hhline}
\usepackage{arydshln}

\begin{document}

\title{Distributed Optimal Generation and Load-Side Control for Frequency Regulation in Power Systems}

\author{Luwei~Yang, Tao~Liu,~\IEEEmembership{Member,~IEEE},~Zhiyuan~Tang,~\IEEEmembership{Student~Member,~IEEE},~and~David~J.~Hill,~\IEEEmembership{Life~Fellow,~IEEE}
\thanks{This work was supported by the Research Grants Council of the Hong Kong Special Administrative Region under the Theme-Based Research Scheme through Project No. T23-701/14-N and General Research Fund Through Project No. 17256516. Liu's work was also supported by the University of Hong Kong Research Committee Research Assistant Professor Scheme.}
\thanks{L. Yang, T. Liu, and Z. Tang are with the Department of Electrical and Electronic Engineering, The University of Hong Kong, Hong Kong S. A. R.,  China (e-mail: lwyang@eee.hku.hk; taoliu@eee.hku.hk; zytan@eee.hku.hk).}
\thanks{D. J. Hill is with the Department of Electrical and Electronic Engineering, The University of Hong Kong, Hong Kong S. A. R., China, and also with the School of Electrical and Information Engineering, The University of Sydney, Sydney, NSW 2016, Australia (e-mail: dhill@eee.hku.hk; david.hill@sydney.edu.au).}}

\maketitle

\begin{abstract}
In order to deal with issues caused by the increasing penetration of renewable resources in power systems, this paper proposes a novel distributed frequency control algorithm for each generating unit and controllable load in a transmission network to replace the conventional automatic generation control (AGC). The targets of the proposed control algorithm are twofold. First, it is to restore the nominal frequency and scheduled net inter-area power exchanges after an active power mismatch between generation and demand. Second, it is to optimally coordinate the active powers of all controllable units in a distributed manner. The designed controller only relies on local information,~computation, and peer-to-peer communication between cyber-connected buses, and it is also robust against uncertain system parameters. Asymptotic stability of the closed-loop system under the designed algorithm is analysed by using a nonlinear structure-preserving model including the first-order turbine-governor dynamics. Final-ly, case studies validate the effectiveness of the proposed method.
\end{abstract}


\IEEEpeerreviewmaketitle
\section{Introduction}
A large frequency excursion caused by active power imbalance between supply and demand may damage devices (e.g., synchronous generators), or even trigger cascading failures and blackouts \cite{agc}. Therefore, to maintain the frequency close to its nominal value (50 Hz or 60 Hz) is a critical task for operating a stable power system \cite{dan}. To achieve such a target, traditional power systems adopt a three-layer frequency controller~including droop control, automatic generation control~(AGC)~and economic dispatch (ED), whose key idea is to make generation follow demand \cite{hc}. Before renewable energy resources were introduced into power systems, this traditional generation-side control paradigm worked well as the system power imbalance mainly results from the variations of loads which usually chan-ge relatively slowly and can be predicted with high accuracy \cite{ed}. However, it may be inadequate to regulate the frequency of a system with high penetration of renewable power. For one thing, synchronous generators may not be able to follow the fast fluctuations of renewable resource outputs.  For another thing, many renewable generating units such as wind turbines and solar panels are connected to the system via power electro-nic devices, which do not provide inertia \cite{wk}.  This may reduce the inertia of the entire system and make the power grid more sensitive to disturbances \cite{inertia}. A possible way to solve the above  issues is to use more fast-ramping generators or energy storage devices as spinning reserves, which will definitely increase the operation costs \cite{both5}. Therefore, how to develop a cost-effective way to maintain the system frequency is of great importance.

To alleviate the negative impacts resulted from renewables, load-side control (demand response) has been advocated to participate in frequency regulation, due to the advantages such as instantaneous responsiveness and distributed availability throughout the grid  \cite{cl,clcl,clclcl}. Various load-side frequency control methods ranging from fully decentralized, distributed to centralized structures have been developed for bulk power systems \cite{cc,cl2,both1,cl3,both2,both3,both6,pd1,pd2,pd4,mf2,
mf1,mf3,mf4,both4,bd,cb1} and microgrids \cite{pd3,cb2,cb3,cb4,mg2,mg3} (to name just a few). The basic idea behind these control methods is to formulate the frequency regulation issue with load-side participation as an optimization problem, and then the controller is synthesized by solving the corresponding optimization problem. Among these control methods, the centralized strategies are vulnerable to single points of failure, and the fully decentralized strategies may lose their effectiveness in the presence of frequency~measurement noises \cite{both5}. Distributed frequency control can strike a balance between the centralized and decentralized methods, and thus has received a great deal of attention. 

Currently, there are three main types of distributed frequen-cy control methods. The first type is primal-dual gradient based approach \cite{cl2,both1,cl3,both2,both3,both6,pd1,pd2,pd4,cc}, where the controller is derived by using a partial primal-dual gradient algorithm to solve~the~optimization problem with respect to frequency regulation. A~main drawback of this approach is that the exact values of the~generator damping and load frequency sensitive coefficients which are usually time-varying and unknown in practice \cite{ed} are needed in the designed controllers. This issue is addressed by the second approach, i.e., the intelligent measurement-based approach. Various advanced intelligent controllers, e.g. artificial neural network (ANN) controller \cite{mf2}, fuzzy logic controller \cite{mf1}, and reinforcement learning (RL) controller \cite{mf3}, have been developed for frequency regulation. However,~these~intelligent techniques may cause a heavy computation burden, and stability of the closed-loop system under the controllers are not theoretically guaranteed \cite{mf4}. The third approach is based on a consensus algorithm that asymptotically converges to some identical marginal costs \cite{both4,bd,cb1,cb2,cb3,cb4,pd3,mg2,mg3}. The frequency~control task is formulated as an optimization problem and then is solved via a consensus-based controller. The advantages of these consensus-based approaches are that they are easy~to~im-plement and stability of the closed-loop system can be guaran-teed. However, most of the existing works in this type of ap-proaches did not consider the issue of keeping the scheduled net inter-area power exchanges, which is also a key task for frequency regulation in power systems.

In view of the abovementioned problems, this paper studies the frequency regulation issue of power systems and proposes a fully distributed frequency control algorithm for each generating unit and controllable load in a transmission network to replace the conventional AGC. It proves that the proposed algorithm is able to regulate the system frequency and net tie-line power flows between interconnected control areas with a minimum total operation cost. The contributions of~the~paper with respect to the existing literature are summarized below

(\romannumeral1) Asymptotic stability conditions of the closed-loop system under~the proposed frequency control algorithm are obtained, where the nonlinear structure-preserving model including the typical first-order turbine-governor dynamics is adopted in~the stability analysis. This complements the existing studies, e.g., \cite{cc,cl2,both1,cl3,pd1,pd4,mf2,mf1,mf3}, where the stability analysis is developed only based on a linearized model. Moreover, the turbine-governor dynamics are neglected in \cite{cc,cl2,cl3,pd1,pd4}.

(\romannumeral2) Compared to \cite{both2,both3,pd2,pd1,cl2,pd4,cc,both1,both4,bd,cb1}, our control algorithm is able to restore the scheduled net inter-area power flows after disturbances. In particular, unlike \cite{both6,cl3} where centralized algorithms are designed to fulfil the inter-area flow requirement, the proposed controller is fully distributed and only relies on local information, computation and peer-to-peer communication~between cyber-connected buses.

(\romannumeral3) Different from the distributed frequency controllers proposed in \cite{both6,pd2,cl3,pd1,both1,both4,both2,both3} which require knowledge of the exact values of all generator damping and load frequency sensitive coefficients, our control algorithm is robust against these uncertain system parameters.

The remainder of this paper proceeds as follows. In section \uppercase\expandafter{\romannumeral2}, we introduce the power system model and formulate the frequency regulation issue with load-side participation as an optimization problem. In section \uppercase\expandafter{\romannumeral3},  we present the proposed distributed optimal frequency control method, and analyse the optimality as well as stability of the equilibrium point of the closed-loop system under the proposed controller. In section \uppercase\expandafter{\romannumeral4}, case studies are conducted to verify the effectiveness of the control algorithm. Finally, conclusions are given in section \uppercase\expandafter{\romannumeral5}.

\textit{Notations:} Denote the set of real numbers, $n$-dimensional real vectors, $(m\times n)$-dimensional real matrices by $\mathbb{R}$, $\mathbb{R}^{n}$, and $\mathbb{R}^{m\times n}$, respectively. The notations $\text{diag}(a_1,\dots,a_k)$ and $\text{diag}(A_1,\dots,A_k)$ represent the diagonal and block diagonal matrices with $a_{i}\in\mathbb{R}$ and $A_{i}\in\mathbb{R}^{m_i\times n_i}$, $i=1,\dots,k$, respectively. Let $\text{col}(x_1,\dots,x_k)= (x_1^T,\dots,x_k^T)^T$ denote the column vector consisting of vectors $x_i\in\mathbb{R}^{n_i}$, $i=1,\cdots,k$. Denote $1_n$ as the $n$-dimensional vector with all entries equal to $1$, $I_{n}$ as the $n$-dimensional identity matrix,  $0_{n\times n}$ as the $(n\times n)$-dimensional zero matrix. In this paper, we will drop the subscripts of vectors and matrices when they are obvious in the context. For a function $f:\mathbb{R}^n\to\mathbb{R}$, we use $(\nabla f)^{-1}(\cdot)$ to denote the inverse of its gradient $\nabla f$ if $\nabla f$ is invertible.

\section{Problem Formulation}

We consider a power transmission network with $n$ buses, $l$ transmission lines, and $k$ control areas, whose index sets are defined by $\mathcal{N}=\{1,\dots,n\}$, $\mathcal{L}_p=\{1,\dots,l\}$, and $\mathcal{K}=\{1,\dots,k\}$, respectively. We use the undirected graph $G_p(\mathcal{N},\mathcal{L}_p)$ to represent the topology of the transmission network, and interchangeably use $e$ and $(i,j)$ to denote an edge in the graph that connects buses $i$ and $j$. By assigning an arbitrary orientation to each edge $e\in\mathcal{L}_p$, the incidence matrix of $G_{p}(\mathcal{N},\mathcal{L}_p)$ can be defined as $C_p=[C_{ie}]\in\mathbb{R}^{n\times l}$, where $C_{ie}=1$, if bus $i$ is the source of $e$, $C_{ie}=-1$, if bus $i$ is the end of $e$, and $C_{ie}=0$ otherwise. 

In this paper, we adopt the following standard assumptions that are extensively used in transmission networks \cite{agc}.

(\romannumeral1) The transmission network is connected and lossless;

(\romannumeral2) The frequency is mainly affected by active power flows, and the impacts from reactive power flows are ignored;

 (\romannumeral3) Bus voltage magnitudes $\vert V_i\vert$, $i\in\mathcal{N}$, are fixed.\\
In fact, these assumptions are generally valid in real-world transmission networks \cite{wk}.

To describe the dynamics of the power network, we use the nonlinear structure-preserving model proposed in \cite{sp}. We partition the buses into $n_g$ generator buses and $n_l$ load buses, and define the corresponding index sets as $\mathcal{{N_G}}=\{1,\dots,n_g\}$ and $\mathcal{N_L}=\{n_g+1,\dots,n\}$. Hence, we have $n=n_g+n_l$, and $\mathcal{N}=\mathcal{N_G}\cup\mathcal{N_L}$. Further, we assume that each load bus has an aggregate controllable load, and consider the first-order turbine-governor dynamics for generators. For each bus $i\in\mathcal{N}$, let $\omega_i$ be the frequency deviation from the nominal value; $P_{e}=$ $T_{p_{e}}\text{sin}(C_{ie}\theta_i+C_{je}\theta_j)$ be the power flow along transmission line $e\in\mathcal{L}_p$ with $T_{p_{e}}=\vert V_{i}\vert\vert V_{j}\vert Y_{ij}$ and $Y_{ij}$ being the suscep-tance of  $e$. For each generator bus $i\in\mathcal{N_G}$, let $\theta_i$ be the power angle with respect to a synchronously rotating reference; $P_{m_i}$, $P_{c_i}$ be the mechanical power input and load reference set-point, respectively; $M_i$, $T_{i}$, $R_i$ represent the rotational inertia, turbine-governor time constant and droop~control~gain,~respectively; $D_i>0$ be the damping coefficient. For each load bus $i\in\mathcal{N_L}$, let $\theta_i$ be the voltage phase angle; $r_i$, $d_i$ be the active power consumed by uncontrollable load and controllable load, respectively; $D_i$ be the load frequency sensitive coefficient. Here, we assume that all load buses satisfy $D_i>0$, $i\in\mathcal{N_L}$.

Define $\theta=\text{col}(\theta_1,\dots,\theta_n)$, $\omega_{\mathcal{G}}=\text{col}(\omega_1,\dots,\omega_{n_g})$, $\omega_{\mathcal{L}}=\text{col}(\omega_{n_g+1},\dots, \omega_n)$, $\omega=\text{col}(\omega_{{\mathcal{G}}},\omega_{{\mathcal{L}}})$, $P_{m}=\text{col}(P_{m_1},\dots,$ $P_{m_{n_g}})$, $P_{c}=\text{col}(P_{c_1},\dots,P_{c_{n_g}})$, $r=\text{col}(r_{n_g+1},\dots,r_n)$, $d=\text{col}(d_{n_g+1},\dots,d_{n})$, and $P=\text{col}(P_1,\dots,P_l)$. Then, the mathematical model of the power system is given as follows
\begin{equation}
\label{model}
\begin{split}
\dot{\theta}&=\omega\\
M_\mathcal{G}\dot{\omega}_{\mathcal{G}}&=-D_{\mathcal{G}}\omega_{\mathcal{G}}+P_{m}-C_{p_{\mathcal{G}}}P\\
T\dot{P}_{m}&=-R^{-1}\omega_{\mathcal{G}}-P_{m}+P_{c}\\
0&=-D_{\mathcal{L}}\omega_{\mathcal{L}}-d-r-C_{p_{\mathcal{L}}}P\\
P&=T_p\text{sin}(C_p^T\theta)
\end{split}
\end{equation}
where $M_\mathcal{G}=\text{diag}(M_1,\dots,M_{n_g})$,  $D_{\mathcal{G}}=\text{diag}(D_{1},\dots,$ $D_{n_g})$, $D_{\mathcal{L}}=\text{diag}(D_{n_g+1},\dots,D_{n})$, $T=\text{diag}(T_{1},\dots,T_{n_g})$, $R=\text{diag}(R_{1},\dots,R_{n_g})$, and $T_p=\text{diag}(T_{p_{1}},\dots,T_{p_{l}})$. Matrices $C_{p_\mathcal{G}}$ and $C_{p_\mathcal{L}}$ are the submatrices of $C_p$, and are derived by collecting the rows of $C_p$ indexed by ${\mathcal{N_G}}$ and ${\mathcal{N_L}}$, respectively.

\textit{Remark 1:} It should be pointed out that model \eqref{model} can also describe the dynamics of inverter-connected renewable generating units,  which can be regarded as negative loads by adding a new term $P_{r_i}$ that is the renewable power generation. The uncontrollable and controllable loads $r_i$ and $d_i$ can be zero or non-zero depending on whether a local load is connected to the renewable generator bus or not \cite{wk}.

The control objective of this paper is to develop a fully distributed optimal frequency control algorithm for system \eqref{model}, which is able to restore the nominal frequency and~scheduled net inter-area power exchanges after disturbances by optimally allocating the active powers of all generating units and controllable loads. To achieve these targets, we denote $F_i(P_{m_i})$ as the generation cost of each generator bus $i\in\mathcal{N_G}$, and $U_i(d_{i})$ as the user utility of each load bus $i\in\mathcal{N_L}$. We further make the following assumptions for these cost/utility functions which are extensively adopted for distributed frequency regulation in power systems (e.g., \cite{both1,cl3,both6})

\textit{Assumption 1:} Functions $F_i(P_{m_i})$, $U_i(d_{i})$ are respectively strongly convex and strongly concave, and are both second-order continuously differentiable with $\nabla^2 F_i(P_{m_i})\geq a_i>0$, $\forall i\in\mathcal{N_G}$, and $\nabla^2 U_i(d_{i})\leq a_i<0$, $\forall i\in\mathcal{N_L}$.

\textit{Assumption 2:} Functions $\nabla F_i(P_{m_i})$, $\nabla U_i(d_{i})$ are Lipschitz continuous with a Lipschitz constant $b_i\geq\vert a_i\vert$, $\forall i\in\mathcal{N}$.\\
Under Assumption 1, $\nabla F_i(P_{m_i})$, $\nabla U_i(d_{i})$ are strictly monotone, and thereby invertible \cite{cp}.

Define $F(P_{m})=\sum_{i\in\mathcal{N_G}}F_{i}(P_{m_{i}})$, $U(d)=\sum\nolimits_{i\in\mathcal{N_L}} U_{i}(d_i)$ as the total generation cost and total user utility of system \eqref{model}, then a controller is said to achieve an optimal power~allocation if it makes the trajectory of system \eqref{model} asymptotically converge to the optimal solution of the following optimal load frequency control (OLFC) problem \cite{cl3,both6}
\begin{subequations}
\label{1}
\begin{align}
\mathop{\text{minimize}}~~~~~~&F(P_m)-U(d)\nonumber\\
\text{subject to}~~~~~~&P_{m}-C_{p_\mathcal{G}}P=0\label{op1}\\
&r+d+C_{p_\mathcal{L}}P=0\label{op2}\\
&EC_pP=P_{t}\label{op3}
\end{align}
\end{subequations}
where matrix $E=\left[E_{si}\right]\in\mathbb{R}^{k\times n}$ is defined as $E_{si}=1$, if $i\in\mathcal{N}_s$, and $E_{si}=0$ otherwise. Here, $\mathcal{N}_s$ is the index set of buses within control area $s\in\mathcal{K}$. $P_{t}=\text{col}(P_{t_1},\dots,P_{t_k})$~consists of the scheduled net tie-line power $P_{t_s}$ of each control area.

In the OLFC problem \eqref{1}, constraints \eqref{op1} and \eqref{op2} represent that the total controllable power increment has to equal to the total net demand change, i.e., $1_{n_g}^TP_{m}=1_{n_l}^Td+1_{n_l}^Tr$. Constraint \eqref{op3} is to preserve the scheduled net power interchanges~between physically interconnected control areas. We assume the OLFC problem \eqref{1} is feasible. Then, its optimality conditions can be determined by using the Karush-Kuhn-Tucker (KKT) conditions \cite{cp}, and are summarized in the following lemma

\lemma\label{lem1} The feasible solution col($\bar{P}_m,\bar{d},\bar{P}$) of OLFC \eqref{1} is optimal if and only if there exist constants $\bar{\lambda}_i$, $i\in\mathcal{N}$, and $\bar{\Lambda}_s$, $s\in\mathcal{K}$, satisfying
\begin{subequations}
\label{kkt}
\begin{align}
\nabla F(\bar{P}_{m})+\bar{\lambda}_{\mathcal{G}}&=0\label{kkt1}\\
\nabla U(\bar{d})+\bar{\lambda}_{\mathcal{L}}&=0\label{kkt2}\\
\bar{\lambda}-E^T\bar{\Lambda}&=0\label{kkt3}
\end{align}
\end{subequations}
where $\bar{\lambda}_\mathcal{G}=\text{col}(\bar{\lambda}_1,\dots,\bar{\lambda}_{n_g})$, $\bar{\lambda}_\mathcal{L}=\text{col}(\bar{\lambda}_{n_g+1},\dots,\bar{\lambda}_{n})$, $\bar{\lambda}=\text{col}(\bar{\lambda}_{\mathcal{G}},\bar{\lambda}_{\mathcal{L}})$, and $\bar{\Lambda}=\text{col}(\bar{\Lambda}_1,\dots,\bar{\Lambda}_k)$.

\textit{Proof:} The Lagrangian function $L=L(P_{m},d,P,\lambda,\mu)$ of the OLFC problem \eqref{1} is given by
\begin{align}
\label{lap}
L=&F(P_{m})-U(d)+{\lambda}^T_{\mathcal{G}}(P_{m}-C_{p_\mathcal{G}}P)-\lambda^T_{\mathcal{L}}(r+d+C_{p_\mathcal{L}}P)\nonumber\\
&+\mu^T(EC_pP-P_t)
\end{align} 
with multipliers $\lambda_{\mathcal{G}}=\text{col}(\lambda_1,\dots,\lambda_{n_g})$, $\lambda_{\mathcal{L}}=\text{col}(\lambda_{n_g+1},\dots,$ $\lambda_{n})$, $\lambda=\text{col}(\lambda_{\mathcal{G}},\lambda_{\mathcal{L}})$, and $\mu=\text{col}(\mu_1,\dots,\mu_k)$. The primal and dual feasibility of KKT conditions implies that the feasible solution $\text{col}(\bar{P}_{m},\bar{d},\bar{P})$ satisfies constraints \eqref{op1}-\eqref{op3}. The stationarity of KKT conditions at the optimality, i.e., $\frac{\partial}{\partial P_{m}}L=0$, $\frac{\partial}{\partial d}L=0$, $\frac{\partial}{\partial P}L=0$, gives \eqref{kkt1}, \eqref{kkt2}, and
\begin{align}
\label{mu}
C_p^TE^T\bar{\mu}-C_p^T\bar{\lambda}=0
\end{align}
where $\bar{\mu}=\text{col}(\bar{\mu}_1,\dots,\bar{\mu}_k)$, and $\bar{\mu}_s$, $s\in\mathcal{K}$ is the value of $\mu_s$ at the optimality. To deduce \eqref{mu}, we use the fact that $C_p^T\bar{\lambda}=C_{p_\mathcal{G}}^T\bar{\lambda}_{\mathcal{G}}+C_{p_\mathcal{L}}^T\bar{\lambda}_{\mathcal{L}}$. Since graph $G_p(\mathcal{N},\mathcal{L}_p)$ is connected and undirected, the null space of matrix $C_p^T$ is $\text{span}(1_{n})$ \cite{im}. Hence, equation \eqref{mu} yields $E^T\bar{\mu}-\bar{\lambda}=\nu 1_{n}$, or equivalently, $E_i^T\bar{\mu}-\bar{\lambda}_i=\nu$ with some $\nu\in\mathbb{R}$, where $E_i\in\mathbb{R}^{k}$ denotes the vector derived by refining the $i$th column of matrix $E$. According to the definition of $E$, $E_i$ is the vector with the $s_i$th entry being one and other entries being zero, where $s_i\in\mathcal{K}$ denotes the index of the control area that bus $i$ belongs to. Then, we have $E_i^T\bar{\mu}=\bar{\mu}_{s_i}$, and thus, $\bar{\lambda}_{i}=\bar{\mu}_{s_i}-\nu$, which implies that $\bar{\lambda}_{i}$ is identical for all buses within the same control area. Without loss of generality, for control area $s$, we let $\bar{\lambda}_i=\bar{\Lambda}_s$, $\forall i\in\mathcal{N}_s$. Then, it follows from the definition of $E$ that $\bar{\lambda}=E^T\bar{\Lambda}$, and hence, the results in Lemma \ref{lem1} follows. $\hfill\blacksquare$

\textit{Remark 2:} The quadratic cost/utility functions of the form $F_i(P_{m_i})=\frac{c_{1i}}{2}P_{m_i}^2+c_{2i}P_{m_i}+c_{3i}$, $U_i(d_{i})=\frac{c_{1i}}{2}d_{i}^2+c_{2i}d_{i}+c_{3i}$, with $c_{1i}>0$, $\forall i\in\mathcal{N_G}$, and $c_{1i}<0$, $\forall i\in\mathcal{N_L}$, which are commonly used to quantify the costs of generators \cite{pd2,both1,bd} as well as utilities of controllable loads \cite{cc,both1,bd}, are special cases of the objective functions adopted in the paper and satisfy Assumptions 1 and 2. In particular, in terms of the quadratic utility function, it has been shown in \cite{qlu} that an end user usually values its power consumption according to a declining marginal benefit as a function of consumed amount. Let the marginal benefit be described by $u_i(d_i)=c_{1i}d_i+c_{2i}$, where $c_{2i}$ is the value of the very first unit of power consumed, and $c_{1i}$ is how rapidly the marginal value of additional
consu-mption declines. Then, the user's utility is the integral of this marginal benefit, which leads to the quadratic form presented above. For more details of physical meanings of the quadratic utility/cost functions, please refer to \cite{qlu}, \cite{qgc}, respectively.

\textit{Remark 3:} The OLFC problem \eqref{1} is designed for multi-area power systems which contain the single-area power systems as special cases. For a multi-area power system, according to Lemma \ref{lem1}, conditions \eqref{kkt1} and \eqref{kkt2} require that the incremental cost/utility value of bus $i\in\mathcal{N}$, i.e., the first derivative of the corresponding cost/utility function, equals to $-\bar{\lambda}_i$ at the optimality. Condition \eqref{kkt3} requires that $\bar{\lambda}_i$ is identical for all buses within the same control area, i.e., the optimal power allocation among all generating units and controllable loads has an identical incremental cost/utility value for buses located in the same control area. This is due to the fact that the inter-area power exchanges are fixed at the scheduled values, and hence, the optimal power allocation of controllable units only occurs inside the control area. In this case, the incremental cost/utility functions for buses that belong to different control areas do not necessarily reach the same values, and the~differences can be regarded as the price of imposing the tie-line bias constraint \eqref{op3}. For a single-area power system, the corresponding optimization problem can be obtained by removing \eqref{op3}. Then, the KKT conditions are reduced into \eqref{kkt1}, \eqref{kkt2} with $
\bar{\lambda}=\nu 1_n,~\nu\in\mathbb{R}$,
which means all buses should have the same incremental cost/utility values at the optimality.

\textit{Remark 4:} In this paper, we do not consider the capacity constraints on each generating unit and controllable load. However, in practice, each controllable unit can only adjust its power output within a certain range, i.e., $P_{m_i}\in[P_{m_i}^{\text{min}},P_{m_i}^\text{max}]$, $i\in\mathcal{N_G}$, and $d_i\in[d_{i}^\text{min},d_i^{\text{max}}]$, $i\in\mathcal{N_L}$. For this case, the distributed projection-based control method proposed in  \cite{vp} can be used to solve the problem. Then, the problem becomes more complicated and will be studied in the future.

\section{Distributed Optimal Frequency Regulation}

To achieve the control objectives formulated above, we will design a fully distributed control algorithm to coordinate the controllable units in system \eqref{model}, and analyse stability of the closed-loop system under the designed algorithm.

\subsection{Distributed Control Algorithm}

We assign each control area $s\in\mathcal{K}$ a connected and undirected communication network, and use graph $G_{c_s}(\mathcal{N}_{s},\mathcal{L}_{c_s})$ to represent its topology which can be different from the physical transmission network, where $\mathcal{L}_{c_s}$ denotes the communication link set of the $s$th control area. Furthermore, we connect two different graphs $G_{c_s}(\mathcal{N}_{s},\mathcal{L}_{c_s})$ and $G_{c_{\bar{s}}}(\mathcal{N}_{\bar{s}},\mathcal{L}_{c_{\bar{s}}})$ by adding communication links if the two control areas $s,\bar{s}\in\mathcal{K}$ are physically interconnected in the grid. The new added~communication links have the same ends as the~corresponding tie lines. The distributed algorithm to be designed will rely on information exchanges between these cyber-connected buses.

We use $G_c(\mathcal{N},\mathcal{L}_c)$ to denote the topology of the entire communication network for the power system, where $\mathcal{L}_c=\mathcal{L}_{c_1}\cup\dots\cup\mathcal{L}_{c_k}\cup\mathcal{L}_b$ with set $\mathcal{L}_b$ consisting of the communication edges that connect different control areas. We assume that any two cyber-connected buses in $G_c(\mathcal{N},\mathcal{L}_c)$ can get access to each other's information via bidirectional communication. We define $L_c=[l_{c_{ij}}]\in\mathbb{R}^{n\times n}$ as the Laplacian matrix of graph $G_c(\mathcal{N},\mathcal{L}_c)$, where $l_{c_{ij}}=l_{c_{ji}}<0$ indicates a communication link with weight $-l_{c_{ij}}$ between buses $i$ and $j$, i.e.,  $(i,j)\in\mathcal{L}_c$, $l_{c_{ij}}=0$ indicates no direct connection between buses $i$ and $j$, and $l_{c_{ii}}=-\sum_{i\neq j}l_{c_{ij}}$. Moreover, we denote $G_c(\mathcal{N},\mathcal{L})$ as the subgraph of $G_c(\mathcal{N},\mathcal{L}_c)$ by deleting all edges $(i,j)\in\mathcal{L}_b$ in $\mathcal{L}_c$, i.e., $\mathcal{L}=\mathcal{L}_{c_1}\cup\dots\cup\mathcal{L}_{c_k}$. Then, the Laplacian matrix $L=[l_{{ij}}]\in\mathbb{R}^{n\times n}$ of $G_c(\mathcal{N},\mathcal{L})$ can be defined in a similar way as $L_c$, i.e., $l_{ij}=l_{c_{ij}}$ if $(i,j)\in\mathcal{L}$, $l_{{ii}}=-\sum_{i\neq j}l_{{ij}}$, and $l_{ij}=0$ otherwise. According to the Lemma 4.3 in \cite{im}, the null space of the Laplacian matrices $L_c$, $L$ are $\text{span}(1_n)$ and $\text{span}(E^T)$, respectively, as the undirected graph $G_c(\mathcal{N},\mathcal{L}_c)$ is connected and $G_c(\mathcal{N},\mathcal{L})$ has $k$ connected components.

Now, we present the designed control algorithm. For each bus $i\in\mathcal{N}$, the distributed controller is given as follows
\begin{equation}
\label{ca}
\begin{split}
\dot{P}_{c_i}=&P_{m_i}-(1+\alpha^2_ia_i)P_{c_i}\\
&+\alpha^2_ia_i(\nabla F_i)^{-1}(-\lambda_i-\alpha_i^{-1}M_{i}\omega_i),~i\in\mathcal{N_G}\\
\dot{d}_{i}=&\alpha^2_ia_id_i-\alpha^2_ia_i(\nabla U_i)^{-1}(-\lambda_i)+\omega_i,~i\in\mathcal{N_L}\\
\dot{\lambda}_{i}=&\alpha_i^{-1}K_{i}\omega_i-\alpha_i^{-1}P_{m_i}-R_i P_{c_i}\\
&+(\alpha_i^{-1}+R_i)(\nabla F_i)^{-1}(-\lambda_i-\alpha_i^{-1}M_i\omega_i)\\
&+\alpha_i^{-1}\sum\nolimits_{(i,j)\in\mathcal{L}_{c}}l_{c_{ij}}(\phi_{i}-\phi_j)\\
&+\alpha_i^{-1}\sum\nolimits_{e\in\mathcal{L}_p}C_{ie}P_{e},~i\in\mathcal{N_G}\\
\dot{\lambda}_{i}=&\alpha_i^{-1}K_i\omega_{i}+(\alpha_i^{-1}+1)d_{i}\\
&-(\alpha_i^{-1}+1)(\nabla U_i)^{-1}(-\lambda_i)\\
&+\alpha_i^{-1}\sum\nolimits_{(i,j)\in\mathcal{L}_c}l_{c_{ij}}(\phi_{i}-\phi_j)\\
&+\alpha_i^{-1}\sum\nolimits_{e\in\mathcal{L}_{p}}C_{ie}P_{e},~i\in\mathcal{N_L}\\
\dot{\phi}_{i}=&-\sum\nolimits_{(i,j)\in\mathcal{L}_c}l_{c_{ij}}(M_i\omega_i-M_j\omega_j)\\
&-\sum\nolimits_{(i,j)\in\mathcal{L}_c}l_{c_{ij}}(\alpha_i\lambda_{i}-\alpha_j\lambda_j)\\
&+\sum\nolimits_{(i,j)\in\mathcal{L}_c}l_{c_{ij}}(\gamma_i-\gamma_j),~i\in\mathcal{N}\\
\dot{\gamma}_{i}=&\sum\nolimits_{(i,j)\in\mathcal{L}}l_{{ij}}(z_i-z_j)+\sum\nolimits_{(i,j)\in\mathcal{L}}l_{{ij}}(\gamma_i-\gamma_j)\\
&-\sum\nolimits_{(i,j)\in\mathcal{L}_c}l_{c_{ij}}(\phi_i-\phi_j)-J_{i}^TP_{t},~i\in\mathcal{N}\\
\dot{z}_{i}=&-\sum\nolimits_{(i,j)\in\mathcal{L}}l_{{ij}}(\gamma_i-\gamma_j),~i\in\mathcal{N}
\end{split}
\end{equation}
where $\lambda_i$, $\phi_i$, $\gamma_i$, $z_i$ are four auxiliary variables; $a_i\in\mathbb{R}$ is the constant defined in Assumption 1; $M_i=0$ for all load buses $i\in\mathcal{N_L}$; $(\nabla F_i)^{-1}(\cdot)$, $(\nabla U_i)^{-1}(\cdot)$ are the inverse functions of the gradients of $F_i(P_{m_i})$ and $U_i(d_i)$, respectively; $\alpha_i>0,$ $K_i\geq 0$, $i\in\mathcal{N},$ are control gains to be designed. Particularly, $\alpha_i$ is identical for all buses within the same control area $s\in\mathcal{K}$, i.e., $\alpha_i=\alpha_j$, $\forall i,j\in\mathcal{N}_s$. In \eqref{ca}, we assume that the scheduled net tie-line power $P_{t_s}$ is only known to one bus $i_s\in\mathcal{N}_s$ that lies in the $s$th control area (bus $i_s$ can~be arbitrarily selected). Hence, $J_i\in\mathbb{R}^{k}$ is a vector with the $s$th entry being 1 and other entries being zero if $i= i_s$, and a zero vector if $i\neq i_s$.

In order to achieve the control targets in a distributed way, we introduce four auxiliary variables $\lambda_i$, $\phi_i$, $\gamma_i$ and $z_i$ in \eqref{ca}, where $\lambda_i$, $\phi_i$ are designed to track the incremental cost/utility value, phase angle of bus $i\in\mathcal{N}$, and $\gamma_i$, $z_i$ are~introduced~to ensure that the incremental cost/utility value of each bus and net tie-line power of each area satisfy~the feasibility condition \eqref{op3} and optimality condition \eqref{kkt3} of the OLFC problem. Thus, variables  $\lambda_i$, $\phi_i$ can be interpreted as the virtual incremental cost/utility and virtual phase angle at bus $i$, respectively. In fact, we will show later that $\lambda_i=-\nabla F_i(\bar P_{m_i})$, $\forall i\in\mathcal{N_G}$, $\lambda_i=-\nabla U_i(\bar d_{i})$, $\forall i\in\mathcal{N_L}$, and $-\sum_{(i,j)\in\mathcal{L}_c}l_{c_{ij}}(\phi_i-\phi_j)=\sum_{e\in\mathcal{L}_p}C_{ie}\bar{P}_{e}$, $\forall i\in\mathcal{N}$ at the steady state, where $\nabla F_i(\bar P_{m_i})$, $\nabla U_i(\bar d_{i})$ are~the~optimal~incremental cost, utility values of bus $i$, respectively; and the term $\sum_{e\in\mathcal{L}_p}C_{ie}\bar{P}_{e}$ is the optimal net load flow at bus $i$, and thus the term $-\sum_{(i,j)\in\mathcal{L}_c}l_{c_{ij}}(\phi_i-\phi_j)$, $i\in\mathcal{N}$, can be considered as  the virtual net load flow at bus $i$ in terms of the DC power flow model \cite{agc}.  

To make sure that the designed distributed controller \eqref{ca} achieves the optimal incremental cost/utility value and satisfies the tie-line power bias constraints, we introduce the auxiliary variables $\gamma_i$ and $z_i$. Particularly, $\gamma_i$ is to guarantee that the incremental cost/utility value of each bus satisfies the optimality condition \eqref{kkt3}, and $z_i$ is to guarantee the net tie-line power of each control area equals to its scheduled value. Specifically, we will show later that $\lambda_i=\alpha_i^{-1}\gamma_i+\alpha_i^{-1}{\delta}$ and $\gamma_i=\alpha_i E_{i}^T\bar\Lambda-\delta$  with some $\delta\in\mathbb{R}$ at the equilibrium point. Thus, $\gamma_i$ forces $\lambda_i=E_{i}^T\bar\Lambda$ at the equilibrium point, which satisfies condition \eqref{kkt3}. Additionally, system \eqref{model} under controller~\eqref{ca}~satisfies $-\sum_{(i,j)\in\mathcal{L}}l_{ij}(z_i-z_j)=\sum_{e\in\mathcal{L}_p}C_{ie}P_{e}-J_iP_t$, $\forall i\in\mathcal{N}$,  at the equilibrium point. According to the definition of vector $J_i$, for buses that do not know the scheduled tie-line power, the term $-\sum_{(i,j)\in\mathcal{L}}l_{ij}(z_i-z_j)$ at the steady state is actually the net load flow at bus $i$; and for buses that know the scheduled tie-line power, the term $-\sum_{(i,j)\in\mathcal{L}}l_{ij}(z_i-z_j)$ at the steady state represents the deviation of the net load flow at bus $i$ from the scheduled net tie-line power of the control area that bus $i$ belongs to. Summing $-\sum_{(i,j)\in\mathcal{L}}l_{ij}(z_i-z_j)$ of all buses in the same control area gives 
\begin{align}\label{z}
-\sum_{i\in\mathcal{N}_s}\sum_{(i,j)\in\mathcal{L}}l_{ij}(z_i-z_j)=\sum_{i\in\mathcal{N}_s}\sum_{e\in\mathcal{L}_p}C_{ie}P_{e}-P_{t_s},
\end{align}
$\forall s\in\mathcal{K}$ at the steady state. Based on the definition of Laplacian matrix $L$, the left-hand side of equation \eqref{z} equals to zero, and the right-hand side of \eqref{z} is actually the difference between the actual and scheduled net tie-line power of the $s$th control area. Hence, the introduction of $z_i$ forces each control area to preserve the scheduled net tie-line power at the steady state.

We now illustrate how the designed control algorithm \eqref{ca} works. The auxiliary variables $\lambda_i$, $\phi_i$, $\gamma_i$, $z_i$, and control commands $P_{c_i}$, $i\in\mathcal{N_G}$, $d_i$, $i\in\mathcal{N_L}$, are computed by each bus $i\in\mathcal{N}$ in real time based on local information and information received from the neighbouring buses. Then, each generating unit and controllable load evolve according to their related control commands $P_{c_i}$ and $d_i$, respectively.~Here, it is worth pointing out, to proceed the control processes, bus $i$ requires $M_j\omega_j$, $\lambda_j$, $\phi_j$, $\gamma_j$ from all of its cyber-connected buses $j$, i.e., $(i,j)\in\mathcal{L}_c$, but requires  $z_j$ only from the cyber-connected buses in the same control area, i.e., $(i,j)\in\mathcal{L}$.

\textit{Remark 5:} Most of the existing results on frequency regulation (e.g., \cite{ed,cl3,both6}) adopt centralized algorithms to achieve the scheduled net tie-line power interchange constraint \eqref{op3}, where a control center  is assigned to each control area to gather (broadcast) information from (to) the corresponding buses. As mentioned in the Introduction, such a centralized control method is vulnerable to single-point failures. To overcome this issue, the designed controller \eqref{ca} is fully distributed, and only relies on local information, computation, and peer-to-peer communication between cyber-connected buses.

\textit{Remark 6:} In our control algorithm, both the topology of the communication network $G_{c_{s}}(\mathcal{N}_s,\mathcal{L}_{c_s})$ for each control area $s\in\mathcal{K}$ and the constant weight $-l_{c_{e}}$ for each communication link $e\in\mathcal{L}_c$ can be arbitrarily selected, which will not affect the system equilibrium point and its stability as long as $G_{c_{s}}(\mathcal{N}_s,\mathcal{L}_{c_s})$ is connected and $l_{c_{e}}<0$ for any $e\in\mathcal{L}_c$ (see Theorem \ref{the1} and Theorem \ref{the2} for details). However, different network topology or weights may have significant impacts on the system transient performance (e.g. frequency nadir and convergence rate). Thus, how to select an optimal topology with appropriate weights for the communication network should be studied in the future.

\subsection{Optimality}

In this subsection, we will show that the equilibrium point of system \eqref{model} with the developed control algorithm \eqref{ca} yields an optimal solution of the OLFC problem \eqref{1}. Since the branch power flow $P_e$ is determined by the angle difference between buses $i$ and $j$ that are directly physically interconnected by~the transmission line $e\in\mathcal{L}_p$, we define $\xi_e=C_{ie}\theta_i+C_{je}\theta_j$ as the angle difference across branch $e$. Let $\xi=\text{col}(\xi_1,\dots,\xi_l)$, $\lambda_\mathcal{G}=\text{col}(\lambda_1,\dots,\lambda_{n_g})$, $\lambda_{\mathcal{L}}=\text{col}(\lambda_{n_g+1},\dots,\lambda_{n})$, $\lambda=\text{col}(\lambda_\mathcal{G},\lambda_\mathcal{L})$, $\phi_{\mathcal{G}}=\text{col}(\phi_1,\dots,\phi_{n_g})$, $\phi_{\mathcal{L}}=\text{col}(\phi_{n_g+1},\dots,$ $\phi_n)$, $\phi=\text{col}(\phi_\mathcal{G},\phi_\mathcal{L})$, $\gamma=\text{col}(\gamma_1,\dots,\gamma_{n})$, $z=\text{col}(z_1,\dots,$ $z_n)$. Then, system \eqref{model} under controller \eqref{ca} can be rewritten in the vectorized formulation as follows
\begin{equation}\label{cs}
\begin{split}
\dot{\xi}=&C_p^T\omega\\
M_\mathcal{G}\dot{\omega}_{\mathcal{G}}=&-D_{\mathcal{G}}\omega_{\mathcal{G}}+P_{m}-C_{p_{\mathcal{G}}}T_p\text{sin}(\xi)\\
0=&-D_{\mathcal{L}}\omega_{\mathcal{L}}-d-r-C_{p_{\mathcal{L}}}T_p\text{sin}(\xi)\\
T\dot{P}_{m}=&-R^{-1}\omega_{\mathcal{G}}-P_{m}+P_{c}\\
\dot{P}_{c}=&P_{m}-(I_{n_g}+\alpha^2_\mathcal{G}A_\mathcal{G})P_c\\
&+\alpha^2_\mathcal{G}A_\mathcal{G}(\nabla F)^{-1}(-\lambda_\mathcal{G}-\alpha_\mathcal{G}^{-1}M_{\mathcal{G}}\omega_{\mathcal{G}})\\
\dot{d}=&\alpha^2_{\mathcal{L}}A_\mathcal{L}d-\alpha^2_\mathcal{L}A_\mathcal{L}(\nabla U)^{-1}(-\lambda_\mathcal{L})+\omega_\mathcal{L}\\
\alpha_\mathcal{G}\dot{\lambda}_{\mathcal{G}}=&K_\mathcal{G}\omega_{\mathcal{G}}-P_m-\alpha_\mathcal{G}RP_c\\
&+(I_{n_g}+\alpha_\mathcal{G}R)(\nabla F)^{-1}(-\lambda_\mathcal{G}-\alpha_\mathcal{G}^{-1}M_\mathcal{G}\omega_\mathcal{G})\\
&+C_{p_\mathcal{G}}T_p\text{sin}(\xi)-L_{c_\mathcal{G}}\phi\\
\alpha_\mathcal{L}\dot{\lambda}_{\mathcal{L}}=&K_\mathcal{L}\omega_{\mathcal{L}}+(I_{n_l}+\alpha_\mathcal{L})d\\
&-(I_{n_l}+\alpha_\mathcal{L})(\nabla U)^{-1}(-\lambda_{\mathcal{L}})\\
&+C_{p_\mathcal{L}}T_p\text{sin}(\xi)-L_{c_\mathcal{L}}\phi\\
\dot{\phi}=&L_cM\omega+L_c\alpha\lambda-L_c\gamma\\
\dot{\gamma}=&-Lz-L\gamma+L_c\phi-JP_{t}\\
\dot{z}=&L\gamma
\end{split}
\end{equation}
where $A_{\mathcal{G}}=\text{diag}(a_1,\dots,a_{n_g})$, $A_{\mathcal{L}}=\text{diag}(a_{n_g+1},\dots,a_{n})$, $\alpha_{\mathcal{G}}=\text{diag}(\alpha_1,\dots,\alpha_{n_g})$, $\alpha_{\mathcal{L}}=\text{diag}(\alpha_{n_g+1},\dots,\alpha_{n})$, $\alpha=\text{diag}(\alpha_\mathcal{G},\alpha_{\mathcal{L}})$, $K_\mathcal{G}=\text{diag}(K_1,\dots,K_{n_g})$, $K_\mathcal{L}=\text{diag}(K_{n_g+1},\dots,K_n)$, $M=\text{diag}(M_\mathcal{G},M_{\mathcal{L}})$ with $M_\mathcal{L}=0_{n_l\times n_l}$, $(\nabla F)^{-1}(-\lambda_\mathcal{G}-\alpha_\mathcal{G}^{-1}M_\mathcal{G}\omega_\mathcal{G})=\text{col}((\nabla F_1)^{-1}(-\lambda_1-\alpha_1^{-1}M_1\omega_1),\dots,(\nabla F_{n_g})^{-1}(-\lambda_{n_g}-\alpha_{n_g}^{-1}M_{n_g}\omega_{n_g}))$, $(\nabla U)^{-1}$ $(-\lambda_\mathcal{L})=\text{col}((\nabla U_{n_g+1})^{-1}(-\lambda_{n_g+1}),\dots,(\nabla U_{n})^{-1}(-\lambda_{n}))$. $L_{c_\mathcal{G}}$, $L_{c_\mathcal{L}}$ are submatrices of $L_c$, and are derived by collecting the rows of $L_c$ indexed by $\mathcal{N_G}$ and $\mathcal{N_L}$, respectively. Matrix $J\in\mathbb{R}^{n\times k}$ is defined by $J=[J_1,\dots,J_n]^T$, and satisfies
\begin{align}\label{j}
EJ&=I_k,~1_n^TJ=1_k^T
\end{align}
according to the definition of vectors $J_i$, $i\in\mathcal{N}$ and matrix $E$.

Define $x=\text{col}(\xi,\omega,P_m,P_c,d,\lambda,\phi,\gamma,z)$ as the state of system \eqref{cs}, and let $x^*=\text{col}(\xi^*,\omega^*,P_m^*,P_c^*,d^*,\lambda^*,\phi^*,\gamma^*,$ $z^*)$ be an equilibrium point of \eqref{cs}. The following theorem establishes the relationship between the equilibrium point $x^*$ and the optimal solution of the OLFC problem \eqref{1}.

\theorem\label{the1} The equilibrium point $x^*$ of \eqref{cs} satisfies $\omega^*=0$, $P_c^*=P_{m}^*$, $\lambda_\mathcal{G}^*=-\nabla F(P_m^*)$, $\lambda^*_\mathcal{L}=-\nabla U(d^*)$, $L_c\phi^*=C_pP^*$, $\gamma^*=\alpha\lambda^*-\frac{1}{n}1_n^T\alpha\lambda^*1_n+\frac{1}{n}1_n^T\gamma(0) 1_n$, $\gamma^*=E^T\rho$ with some $\rho\in\mathbb{R}^k$, $Lz^*=C_pP^*-JP_{t}$, where $P^*=T_p\text{sin}(\xi^*)$. Moreover, $\text{col}(P_{m}^*,d^*,P^*)$ is the optimal solution to \eqref{1}.

\textit{Proof:} According to \eqref{cs}, we have 
\begin{subequations}
\begin{align}
&-\alpha^2_\mathcal{G}A_\mathcal{G}(T\dot{P}_m-\alpha_\mathcal{G}\dot{\lambda}_\mathcal{G})-(I_{n_g}+\alpha_\mathcal{G}R)(T\dot{P}_m+\dot{P}_c)\nonumber\\
=&\Psi_1\omega_\mathcal{G}+\alpha^2_\mathcal{G}A_\mathcal{G}C_{p_\mathcal{G}}T_p\text{sin}(\xi)-\alpha^2_\mathcal{G}A_\mathcal{G}L_{c_\mathcal{G}}\phi\label{18a}\\
&(I_{n_l}+\alpha_\mathcal{L})\dot{d}-\alpha^3_\mathcal{L}A_\mathcal{L}\dot{\lambda}_\mathcal{L}\nonumber\\
=&\Psi_2\omega_\mathcal{L}-\alpha^2_\mathcal{L}A_\mathcal{L}C_{p_\mathcal{L}}T_p\text{sin}(\xi)+\alpha^2_\mathcal{L}A_\mathcal{L}L_{c_\mathcal{L}}\phi\label{18b}
\end{align}
\end{subequations}
where $\Psi_1$ and $\Psi_2$ are two positive definite diagonal matrices defined by $\Psi_1=\alpha_{\mathcal{G}}+R^{-1}+\alpha^2_\mathcal{G}A_\mathcal{G}K_\mathcal{G}+\alpha^2_\mathcal{G}A_\mathcal{G}R^{-1}$, and $\Psi_2=\alpha_\mathcal{L}+I_{n_l}-\alpha^2_\mathcal{L}A_\mathcal{L}K_\mathcal{L}$, respectively. Since $\dot{P}_{m}=\dot{P}_{c}=\dot{\lambda}_\mathcal{G}=0$, and $\dot{d}=\dot{\lambda}_\mathcal{L}=0$ at the steady state, the following two equations hold
\begin{subequations}
\begin{align}
\alpha^{-2}_\mathcal{G}A^{-1}_\mathcal{G}\Psi_1\omega^*_\mathcal{G}+C_{p_\mathcal{G}}T_p\text{sin}(\xi^*)-L_{c_\mathcal{G}}\phi^*&=0\label{19a}\\
-\alpha_\mathcal{L}^{-2}A_\mathcal{L}^{-1}\Psi_2\omega^*_\mathcal{L}+C_{p_\mathcal{L}}T_p\text{sin}(\xi^*)-L_{c_\mathcal{L}}\phi^*&=0\label{19b}.
\end{align}
\end{subequations}
Left multiplying equations \eqref{19a}, \eqref{19b} with $1_{n_g}^T$,  $1_{n_l}^T$, respectively, and then summing the two equations gives
\begin{align}\label{omega}
1_{n_g}^T\alpha^{-2}_\mathcal{G}A^{-1}_\mathcal{G}\Psi_1\omega_\mathcal{G}^*-1_{n_l}^T\alpha^{-2}_\mathcal{L}A^{-1}_\mathcal{L}\Psi_2\omega_\mathcal{L}^*=0
\end{align}
where we use the facts $1_{n_g}^TC_{p_\mathcal{G}}+1_{n_l}^TC_{p_\mathcal{L}}=1_{n}^TC_p=0$, and $1_{n_g}^TL_{c_\mathcal{G}}+1_{n_l}^TL_{c_\mathcal{L}}=1_{n}^TL_{c}=0$. Further, since the null space of matrix $C_p^T$ is $\text{span}(1_n)$, $\dot{\xi}=0$ at the steady state implies $\omega^*=\beta 1_n$ with some $\beta\in\mathbb{R}$, and thus, $\omega_\mathcal{G}^*=\beta 1_{n_g}$, $\omega_\mathcal{L}^*=\beta 1_{n_l}$. Substituting $\omega_\mathcal{G}^*=\beta 1_{n_g}$, $\omega_\mathcal{L}^*=\beta 1_{n_l}$ into \eqref{omega} gives
\begin{align}
(1_{n_g}^T\alpha^{-2}_\mathcal{G}A^{-1}_\mathcal{G}\Psi_11_{n_g}-1_{n_l}^T\alpha^{-2}_\mathcal{L}A^{-1}_\mathcal{L}\Psi_21_{n_l})\beta=0
\end{align}
which apparently implies $\beta=0$ by recalling the positive definiteness of diagonal matrices $A_\mathcal{G}$, $\alpha_\mathcal{G}$, $\alpha_\mathcal{L}$, $\Psi_1$, $\Psi_2$ and negative definiteness of matrix $A_\mathcal{L}$. Therefore, we have $\omega^*=0$. 

Solving $\dot{\omega}_\mathcal{G}=0$, $\dot{P}_m=0$, $\dot{P}_c=0$, $\dot{d}=0$, $\dot{\lambda}=0$, $\dot{\phi}=0$, $\dot{\gamma}=0$, and $\dot{z}=0$ with $\omega^*=0$ gives
\begin{subequations}
\begin{align}
P_{m}^*&=C_{p_\mathcal{G}}P^*\label{e1}\\
d^*&=-r-C_{p_\mathcal{L}}P^*\label{e2}\\
P_{c}^*&=P_{m}^*\label{e3}\\
\lambda_\mathcal{G}^*&=-\nabla F(P_{m}^*)\label{e4}\\
\lambda_\mathcal{L}^*&=-\nabla U(d^*)\label{e5}\\
L_c\phi^*&=C_pP^*\label{e6}\\
L_c\gamma^*&=L_c\alpha\lambda^*\label{e7}\\
Lz^*&=C_pP^*-JP_{t}\label{e8}\\
L\gamma^*&=0\label{e9}.
\end{align}
\end{subequations} 
Left multiplying \eqref{e8} with matrix $E$ gives
\begin{align}
EC_pP^*-P_{t}=0
\end{align}
where we use the facts $EJ=I_k$ in \eqref{j} and $EL=0$. Moreover, \eqref{e7} yields $\alpha\lambda^*-\gamma^*=\delta 1_n$ with some $\delta\in\mathbb{R}$. By noting that the summation of the scheduled net tie-line power flows of all control areas is equal to zero, i.e., $1_k^TP_t=0$ \cite{tie}, we can specify $1_n^T\dot{\gamma}\equiv 0$, which indicates $1_n^T\gamma^*\equiv 1_n^T\gamma(0)$. Left multiplying $1_n^T$ on both sides of equation $\alpha\lambda^*-\gamma^*=\delta 1_n$ leads to
\begin{align}
1_n^T\alpha\lambda^*-1_n^T\gamma^*=1_n^T\alpha\lambda^*-1_n^T\gamma(0)=n\delta
\end{align}
which gives $\delta=\frac{1}{n}1_n^T\alpha\lambda^*-\frac{1}{n}1_n^T\gamma(0)$, and hence, 
\begin{equation}\label{g}
\gamma^*=\alpha\lambda^*-\delta 1_n=\alpha\lambda^*-\frac{1}{n}1_n^T\alpha\lambda^*1_n+\frac{1}{n}1_n^T\gamma(0) 1_n.
\end{equation}
Since the null space of $L$ is $\text{span}(E^T)$, \eqref{e9} is equivalent to $\gamma^*=E^T\rho$ with some $\rho\in\mathbb{R}^{k}$. Substituting $\gamma^*=E^T\rho$ into \eqref{g} gives
\begin{align}
\lambda^*=\alpha^{-1}(E^T\rho+\frac{1}{n}1_n^T\alpha\lambda^*1_n-\frac{1}{ n}1_n^T\gamma(0)1_n)
\end{align}
or equivalently,
\begin{align}
\lambda_i^*=\frac{1}{\alpha_i}(E_i^T\rho+\frac{1}{n}1_n^T\alpha\lambda^*-\frac{1}{ n}1_n^T\gamma(0)),~i\in\mathcal{N}.
\end{align}
Due to the characteristic of vector $E_i$ defined in the proof of Lemma \ref{lem1} and the fact that $\alpha_i$ is identical for all buses within the same control area, $\lambda_i^*$ is identical for all buses locating in the same control area. Applying the same arguments as that for establishing $\bar{\lambda}=E^T\bar{\Lambda}$ in Lemma \ref{lem1}, we can easily prove that $\lambda^*$ satisfies condition \eqref{kkt3}.

Now, we can claim that $P_{m}^*$, $d^*$, $P^*$, $\lambda^*$ satisfy the feasible conditions \eqref{op1}-\eqref{op3} and optimality conditions \eqref{kkt1}-\eqref{kkt3} of the OLFC problem \eqref{1}, which implies $\text{col}(P_{m}^*,d^*,P^*)$ is the optimal solution. $\hfill\blacksquare$

\subsection{Stability}
Now we are in a position to analyse stability of the~equilibrium point $x^*$ of system \eqref{cs}. It should be noted that the results derived in this subsection are under the following assumption

\textit{Assumption 3:} The vector $\xi^*$ in the equilibrium point $x^*$ satisfies $\vert\xi_{e}^*\vert<\frac{\pi}{2}$, $\forall e\in\mathcal{L}_p$.\\
Assumption 3 is extensively adopted in power system stability analysis and distributed frequency controller design (e.g., \cite{bd,both6,sp}), and is generally fulfilled under normal operating conditions. Under Assumption 3, the following two lemmas can be obtained

\lemma\label{lem2} Suppose Assumption 3 holds and let $\tilde\xi=\xi-\xi^*$ satisfy $\vert\tilde\xi_{e}+2\xi_{e}^*\vert<\pi$, $\forall e\in\mathcal{L}_p$, the function $W(\tilde\xi)$ defined~by
\begin{equation}
W(\tilde\xi)=1_{l}^TT_p\text{cos}(\xi^*)-1_{l}^TT_p\text{cos}(\tilde\xi+\xi^*)-(T_p\text{sin}(\xi^*))^T(\tilde\xi)
\end{equation}
satisfies $W(\tilde\xi)> 0$, $\forall\tilde{\xi}\neq 0$, and $W(\tilde\xi)=0$ if and only if $\tilde{\xi}=0$, i.e., $\xi=\xi^*$.

\textit{Proof:} The proof is based on the strict convexity of function $\mathcal{W}(\xi)=-1_{l}^TT_p\text{cos}(\xi)$ at point $\xi^*$ and Lemma 4 in \cite{bd}.$\hfill\blacksquare$ 

\lemma \label{lem3} Suppose Assumptions 1-3 hold, the equilibrium point $x^*$ of system \eqref{cs} is unique for any given $\gamma(0)\in\mathbb{R}^{n}$.

\begin{figure*}[hb]
\normalsize
\hrulefill
\setcounter{equation}{23}
\begin{subequations}
\label{vd}
\begin{align}
&\dot{V}(\tilde x)\nonumber\\
=&-\tilde\omega_\mathcal{G}^T(D_{\mathcal{G}}\omega_{\mathcal{G}}-P_{m}+C_{p_{\mathcal{G}}}T_p\text{sin}(\xi))-\tilde{P}_m^TR(R^{-1}\omega_\mathcal{G}+P_m-P_c)-\tilde{P}_c^TR((I_{n_g}+\alpha^2_\mathcal{G}A_\mathcal{G})P_c-P_m-\alpha^2_{\mathcal{G}}A_\mathcal{G}(\nabla F)^{-1}(-\lambda_\mathcal{G}\nonumber\\
&-\alpha_\mathcal{G}^{-1}M_\mathcal{G}\omega_\mathcal{G}))+\tilde{d}^T(\alpha^2_\mathcal{L}A_\mathcal{L}d-\alpha^2_\mathcal{L}A_\mathcal{L}(\nabla U)^{-1}(-\lambda_\mathcal{L})+\omega_\mathcal{L})+\tilde{\phi}^T(L_cM\omega+L_c \alpha\lambda-L_c\gamma)-\tilde{\gamma}^T(Lz+L\gamma-L_c\phi+JP_t)\nonumber\\
&+\tilde{z}^TL\gamma+(M_\mathcal{G}\tilde{\omega}_\mathcal{G}+\alpha_\mathcal{G}\tilde{\lambda}_\mathcal{G})^T((K_\mathcal{G}-D_\mathcal{G})\omega_\mathcal{G}-\alpha_\mathcal{G}RP_c+(I_{n_g}+\alpha_\mathcal{G}R)(\nabla F)^{-1}(-\lambda_\mathcal{G}-\alpha_\mathcal{G}^{-1}M_\mathcal{G}\omega_\mathcal{G})-L_{c_\mathcal{G}}\phi)\nonumber\\
&+(\alpha_\mathcal{L}\tilde{\lambda}_\mathcal{L})^T((K_\mathcal{L}-D_\mathcal{L})\omega_\mathcal{L}+\alpha_\mathcal{L}d-r-(I_{n_l}+\alpha_{\mathcal{L}})(\nabla U)^{-1}(-\lambda_\mathcal{L})-L_{c_\mathcal{L}}\phi)+(T_p\text{sin}(\xi)-T_p\text{sin}(\xi^*))^TC_p^T\omega\label{vd1}\\
=&-\tilde\omega_\mathcal{G}^T(D_{\mathcal{G}}\tilde\omega_{\mathcal{G}}-\tilde P_{m}+C_{p_{\mathcal{G}}}T_p(\text{sin}(\xi)-\text{sin}(\xi^*)))-\tilde{P}_m^TR(R^{-1}\tilde\omega_\mathcal{G}+\tilde P_m-\tilde P_c)-\tilde{P}_c^TR((I_{n_g}+\alpha^2_\mathcal{G}A_\mathcal{G})\tilde P_c-\tilde P_m-\alpha^2_\mathcal{G}A_\mathcal{G}\mathcal{F})\nonumber\\
&+\tilde{d}^T(\alpha^2_\mathcal{L}A_\mathcal{L}\tilde d-\alpha^2_\mathcal{L}A_\mathcal{L}\mathcal{U}+\tilde\omega_\mathcal{L})+\tilde{\phi}^T(L_cM\tilde\omega+L_c\alpha\tilde{\lambda}-L_c\tilde\gamma)-\tilde{\gamma}^T(L\tilde z+L\tilde\gamma-L_c\tilde\phi)+\tilde{z}^TL\tilde\gamma\nonumber\\
&+(M_\mathcal{G}\tilde{\omega}_\mathcal{G}+\alpha_\mathcal{G}\tilde{\lambda}_\mathcal{G})^T((K_\mathcal{G}-D_\mathcal{G})\tilde\omega_\mathcal{G}-\alpha_\mathcal{G}R\tilde P_c+(I_{n_g}+\alpha_\mathcal{G}R)\mathcal{F}-L_{c_\mathcal{G}}\tilde\phi)\nonumber\\
&+(\alpha_\mathcal{L}\tilde{\lambda}_\mathcal{L})^T((K_\mathcal{L}-D_\mathcal{L})\tilde\omega_\mathcal{L}+\alpha_\mathcal{L}\tilde d-(I_{n_l}+\alpha_{\mathcal{L}})\mathcal{U}-L_{c_\mathcal{L}}\tilde\phi)+(T_p\text{sin}(\xi)-T_p\text{sin}(\xi^*))^TC_p^T\tilde\omega\label{vd2}\\
=&-\tilde\omega^T_\mathcal{G}D_\mathcal{G}\tilde\omega_\mathcal{G}+(\alpha_\mathcal{G}\tilde{\lambda}_\mathcal{G}+M_\mathcal{G}\tilde\omega_\mathcal{G})^T(K_\mathcal{G}-D_\mathcal{G})\tilde\omega_\mathcal{G}-(\tilde P_{m}-\tilde P_{c})^TR(\tilde P_{m}-\tilde P_{c})-\tilde\gamma^TL\tilde{\gamma}-\tilde{P}_c^T\alpha^2_\mathcal{G}A_\mathcal{G}R\tilde{P}_c+\tilde{P}_c^T\alpha^2_\mathcal{G}A_\mathcal{G}R\mathcal{F}\nonumber\\
&-\tilde{P}_c^T\alpha_\mathcal{G}R(\alpha_\mathcal{G}\tilde\lambda_\mathcal{G}+M_\mathcal{G}\tilde{\omega}_\mathcal{G})+(\alpha_\mathcal{G}\tilde\lambda_\mathcal{G}+M_\mathcal{G}\tilde{\omega}_\mathcal{G})^T(I_{n_g}+\alpha_\mathcal{G}R)\mathcal{F}-\tilde\omega^T_\mathcal{L}D_\mathcal{L}\tilde\omega_\mathcal{L}+(\alpha_\mathcal{L}\tilde{\lambda}_\mathcal{L})^T(K_\mathcal{L}-D_\mathcal{L})\tilde\omega_\mathcal{G}+\tilde{d}^T\alpha^2_\mathcal{L}A_\mathcal{L}\tilde{d}\nonumber\\
&-\tilde{d}^T\alpha^2_\mathcal{L}A_\mathcal{L}\mathcal{U}+\tilde{d}^T\alpha^2_\mathcal{L}\tilde\lambda_\mathcal{L}-(\alpha_\mathcal{L}\tilde{\lambda}_\mathcal{L})^T(I_{n_l}+\alpha_\mathcal{L})\mathcal{U}\label{vd3}
\end{align}
\end{subequations}
\end{figure*}

\textit{Proof:} Since OLFC \eqref{1} is a strictly convex optimization problem, the corresponding optimal solution is unique, which reveals the uniqueness of $P_m^*$, $d^*$ and $P^*$. Then, based on Theorem 1, the uniqueness of $\omega^*$, $P_c^*$, $\lambda^*$ are obvious from $\omega^*=0$, $P_c^*=P_{m}^*$, $\lambda_\mathcal{G}^*=-\nabla F(P_m^*)$, and $\lambda^*_\mathcal{L}=-\nabla U(d^*)$. For any given $\gamma(0)$, $\gamma^*$ is unique due to equation $\gamma^*=\alpha\lambda^*-\frac{1}{n}1_n^T\alpha\lambda^*1_n+\frac{1}{n}1_n^T\gamma(0) 1_n$ and the uniqueness of $\lambda^*$. Under Assumption 3, the uniqueness of $P^*$ is equivalent to the uniqueness of $\xi^*$.

Now, it only remains to demonstrate that $\phi^*$ and $z^*$ are unique. We prove this by contradiction. From Theorem \ref{the1}, the equilibrium point $x^*$ satisfies $L_c\phi^*=C_pP^*$, and  $Lz^*=C_pP^*-JP_t$. Suppose there exist vectors $\hat{\phi}\neq \phi^*$ and $\hat{z}\neq z^*$ such that $L_c\hat{\phi}=C_pP^*$, and $L\hat{z}=C_pP^*-JP_t$. Then, we have $L_c\hat{\phi}=L_c\phi^*$, and $L\hat{z}=Lz^*$. As the null space of matrices $L_c$, $L$ are respectively $\text{span}(1_n)$ and $\text{span}(E^T)$, we can get 
\setcounter{equation}{20}
\begin{subequations}
\begin{align}
\hat{\phi}&=\phi^*+\varphi1_n\label{28a}\\
\hat{z}&=z^*+E^T\varepsilon \label{28b}
\end{align}
\end{subequations}
where $\varphi\in\mathbb{R}$ and $\varepsilon\in\mathbb{R}^k$. Furthermore, we notice that $1_n^T\dot{\phi}\equiv0$, and $E\dot{z}\equiv0$ from \eqref{cs}, which means that $1_n^T\phi(t)\equiv 1_n^T\phi(0)$, and $Ez(t)\equiv Ez(0)$ holds for $\forall t\geq0$. This indicates
\begin{subequations}
\begin{align}
1_n^T\hat{\phi}&=1_n^T\phi^*=1_n^T\phi(0)\label{29a}\\
E\hat{z}&=Ez^*=Ez(0)\label{29b}.
\end{align}
\end{subequations} 
Substituting \eqref{28a} into \eqref{29a}, we can obtain $1_n^T\phi^*+n\varphi=1_n^T\phi^*$, which implies $\varphi$ must equal to zero. Similarly, substituting \eqref{28b} into \eqref{29b} gives $Ez^*$ $+EE^T\varepsilon=Ez^*$. Then, $\varepsilon$ must be a zero vector, since $EE^T\in\mathbb{R}^{k\times k}$ is a positive definite diagonal matrix with its $s$th diagonal entry being the total number of the buses that the $s$th control area contains. Therefore, $\phi^*=\hat{\phi}$, and $z^*=\hat{z}$, which contradict to our assumption. Now, we can claim that $\phi^*$ and $z^*$ are unique, and hence, the results of Lemma \ref{lem3} follows. $\hfill\blacksquare$

We now present our main results of the paper with respect to asymptotic stability of the closed-loop system \eqref{cs} under the designed control algorithm.

\theorem \label{the2} Consider the closed-loop system \eqref{cs}. Suppose that Assumptions 1-3 hold, and $\alpha_i>0,K_i\geq 0$ satisfy
\begin{subequations}
\label{sc}
\begin{align}
\alpha_i&=\alpha^*_s,~\forall i\in\mathcal{N}_s,~\forall s\in\mathcal{K}\label{hi}\\
\alpha^*_s&<\left(\max\limits_{i\in\mathcal{N}_{s}}\{\varrho_{i}\}\right)^{-1},~\forall s\in\mathcal{K}\label{as}
\end{align}
\end{subequations}
where
\begin{align}
\varrho_{i}&=\frac{ b_iK_i^2}{4D_i}-\frac{b_iK_i}{2}+\frac{ b_iD_i}{4}+\frac{b_iR_i}{a_i}-R_i,~i\in\mathcal{N_G}\nonumber\\
\varrho_{i}&=\frac{ b_iK_i^2}{4D_i}-\frac{b_iK_i}{2}+\frac{ b_iD_i}{4}-\frac{b_i}{a_i}-1,~i\in\mathcal{N_L}.\nonumber
\end{align}
Then the equilibrium point $x^*$ is asymptotically stable.
\begin{figure*}[hb]
\normalsize
\hrulefill
\setcounter{equation}{29}
\begin{subequations}
\label{vde}
\begin{align}
&\dot{V}(\tilde{x})\nonumber\\
\leq&\sum\nolimits_{i\in\mathcal{N_G}}(-D_i\tilde{\omega}^2_i-\alpha_i(K_i-D_i)(\nabla F_i(\eta_{i})-\nabla F_{i}(\zeta_{i}))\tilde{\omega}_i-R_i(\tilde{P}_{m_i}-\tilde{P}_{c_i})^2-\alpha^2_ia_iR_i\tilde{P}_{c_i}^2+\alpha^2_ia_iR_i\vert\eta_{i}-\zeta_{i}\vert\vert\tilde{P}_{c_i}\vert\nonumber\\
&+\alpha^2_iR_{i}\vert\nabla F_i(\eta_{i})-\nabla F_{i}(\zeta_{i})\vert\vert\tilde{P}_{c_i}\vert-b_i^{-1}(\alpha_i+\alpha_i^2R_i)(\nabla F_i(\eta_{i})-\nabla F_{i}(\zeta_{i}))^2)\nonumber\\
&+\sum\nolimits_{i\in\mathcal{N_L}}(-D_i\tilde{\omega}^2_i-\alpha_i(K_i-D_i)(\nabla U_i(\eta_{i})-\nabla U_{i}(\zeta_{i}))\tilde{\omega}_i+\alpha^2_ia_i\tilde{d}_i^2-\alpha^2_ia_i\vert\eta_{i}-\zeta_{i}\vert\vert\tilde{d}_i\vert\nonumber\\
&+\alpha^2_i\vert\nabla U_i(\eta_{i})-\nabla U_{i}(\zeta_{i})\vert\vert\tilde{d}_i\vert-b_i^{-1}(\alpha_i+\alpha^2_i)(\nabla U_i(\eta_{i})-\nabla U_{i}(\zeta_{i}))^2)\\
\leq&\sum\nolimits_{i\in\mathcal{N_G}}(-D_i\tilde{\omega}^2_i-\alpha_i(K_i-D_i)(\nabla F_i(\eta_{i})-\nabla F_{i}(\zeta_{i}))\tilde{\omega}_i-(\alpha_ib_i^{-1}+\alpha_i^2b_i^{-1}R_i-\alpha_i^2a_i^{-1}R_i)(\nabla F_i(\eta_{i})-\nabla F_{i}(\zeta_{i}))^2\nonumber\\
&-R_i(\tilde{P}_{m_i}-\tilde{P}_{c_i})^2-\alpha^2_ia_iR_i\tilde{P}_{c_i}^2+2\alpha^2_iR_{i}\vert\nabla F_i(\eta_{i})-\nabla F_{i}(\zeta_{i})\vert\vert\tilde{P}_{c_i}\vert-\alpha_i^2a_i^{-1}R_i(\nabla F_i(\eta_{i})-\nabla F_{i}(\zeta_{i}))^2)\nonumber\\
&+\sum\nolimits_{i\in\mathcal{N_L}}(-D_i\tilde{\omega}^2_i-\alpha_i(K_i-D_i)(\nabla U_i(\eta_{i})-\nabla U_{i}(\zeta_{i}))\tilde{\omega}_i-(\alpha_ib_i^{-1}+\alpha^2_ib_i^{-1}+\alpha_i^2a^{-1}_i)(\nabla U_i(\eta_{i})-\nabla U_{i}(\zeta_{i}))^2\nonumber\\
&+\alpha^2_ia_i\tilde{d}_i^2+2\alpha^2_i\vert\nabla U_i(\eta_{i})-\nabla U_{i}(\zeta_{i})\vert\vert\tilde{d}_i\vert+\alpha^2_ia_i^{-1}(\nabla U_i(\eta_{i})-\nabla U_{i}(\zeta_{i}))^2)\\
=&-\sum\nolimits_{i\in\mathcal{N_G}}(\alpha_i^2a_iR_i(\vert P_{c_i}\vert-a_i^{-1}\vert\tilde\lambda_i+\alpha^{-1}_iM_i\tilde\omega_i\vert)^2+(\tilde{\omega}_i,\nabla F_i(\eta_{i})-\nabla F_i(\zeta_{i}))Q_i(\tilde{\omega}_i,\nabla F_i(\eta_{i})-\nabla F_i(\zeta_{i}))^T+R_i(\tilde{P}_{m_i}\nonumber\\
&-\tilde{P}_{c_i})^2)+\sum\nolimits_{i\in\mathcal{N_L}}(\alpha_i^2a_i(\vert d_{i}\vert+a_i^{-1}\vert\nabla U_i(\eta_{i})-\nabla U_i(\zeta_{i})\vert)^2+(\tilde{\omega}_i,\nabla U_i(\eta_{i})-\nabla U_i(\zeta_{i}))Q_i(\tilde{\omega}_i,\nabla U_i(\eta_{i})-\nabla U_i(\zeta_{i}))^T)\label{vd4}
\end{align}
\end{subequations}
\end{figure*}

\textit{Proof:} Let $\tilde{x}=x-x^*=\text{col}(\tilde\xi,\tilde\omega,\tilde P_m,\tilde P_c,\tilde d,$ $\tilde \lambda,\tilde \phi,\tilde\gamma,\tilde z)$, where $\tilde\omega=\text{col}(\tilde{\omega}_{\mathcal{G}},\tilde{\omega}_{\mathcal{L}})$, and define set $\Omega_1=\{\tilde x~|~\vert\tilde\xi_{e}+2\xi_{e}^*\vert<\pi,~\forall e\in\mathcal{L}_p\}$ (for physical meaning of set $\Omega_1$, please refer to \cite{sp}). Consider the following Lyapunov function candidate
\setcounter{equation}{24}
\begin{align}
V(\tilde x)
=&\frac{1}{2}(\tilde{\omega}_\mathcal{G}^TM_\mathcal{G}\tilde{\omega}_\mathcal{G}
+\tilde{P}_{m}^TRT\tilde{P}_{m}+\tilde{P}_{c}^TR\tilde{P}_{c}+\tilde{d}^T\tilde{d}+\tilde{\phi}^T\tilde{\phi}\nonumber\\
&+\tilde{\gamma}^T\tilde{\gamma}+\tilde{z}^T\tilde{z})+(M\tilde{\omega}+\alpha\tilde{\lambda})^T(M\tilde{\omega}+\alpha\tilde{\lambda})+W(\tilde\xi)\label{V}.
\end{align}
According to Lemma \ref{lem2}, $V(\tilde x)\geq0$ in $\Omega_1$, and $V(\tilde x)=0$ if and only if $\tilde x=0$, i.e., $x=x^*$.

Taking the time derivative of $V(\tilde x)$ along system \eqref{cs} leads to \eqref{vd}, where $\mathcal{F}=(\nabla F)^{-1}(-\tilde{\lambda}_\mathcal{G}-\alpha_\mathcal{G}^{-1}M_\mathcal{G}\tilde{\omega}_\mathcal{G}-{\lambda}^*_\mathcal{G}-\alpha_\mathcal{G}^{-1}M_\mathcal{G}{\omega}^*_\mathcal{G})-(\nabla F)^{-1}(-{\lambda}^*_\mathcal{G}-\alpha_\mathcal{G}^{-1}M_\mathcal{G}{\omega}^*_\mathcal{G})$, $\mathcal{U}=(\nabla U)^{-1}(-\tilde{\lambda}_\mathcal{L}-{\lambda}^*_\mathcal{L})-(\nabla U)^{-1}(-{\lambda}^*_\mathcal{L})$ with the functions $(\nabla F)^{-1}(\cdot)$ and $(\nabla U)^{-1}(\cdot)$ defined in \eqref{cs}. The equality \eqref{vd1} results from system \eqref{cs}, and the equalities \eqref{vd2}, \eqref{vd3} are derived by using the properties of the equilibrium point $x^*$ specified in Theorem \ref{the1}. 

Let $\eta_{i}=(\nabla F_i)^{-1}(-\tilde\lambda_i-\alpha_i^{-1}M_i\tilde\omega_i-\lambda_i^*-\alpha_i^{-1}M_i\omega^*_i)$, $\zeta_{i}=(\nabla F_i)^{-1}(-\lambda_i^*-\alpha_i^{-1}M_i\omega^*_i)$ for $i\in\mathcal{N_G}$, and $\eta_{i}=(\nabla U_i)^{-1}(-\tilde\lambda_i-\lambda_i^*)$, $\zeta_{i}=(\nabla U_i)^{-1}(-\lambda_i^*)$ for $i\in\mathcal{N_L}$. Then, we have
\begin{subequations}
\begin{align}
\tilde\lambda_i+\alpha^{-1}_iM_i\tilde\omega_i&=-(\nabla F_i(\eta_{i})-\nabla F_i(\zeta_{i})),~ i\in\mathcal{N_G}\label{laf}\\
\tilde\lambda_i&=-(\nabla U_i(\eta_{i})-\nabla U_i(\zeta_{i})),~i\in\mathcal{N_L}\label{lau}
\end{align}
\end{subequations}
Consequently, it follows from \eqref{vd3} and the positive semi-definiteness of the Laplacian matrix $L$ that
\begin{align}
&\dot{V}(\tilde{x})\nonumber\\
\leq&\sum\nolimits_{i\in\mathcal{N_G}}(-D_i\tilde{\omega}^2_i-\alpha_i(K_i-D_i)(\nabla F_i(\eta_{i})-\nabla F_{i}(\zeta_{i}))\tilde{\omega}_i\nonumber\\
&-R_i(\tilde{P}_{m_i}-\tilde{P}_{c_i})^2-\alpha^2_i a_iR_i\tilde{P}_{c_i}^2+ \alpha^2_ia_iR_i(\eta_{i}-\zeta_{i})\tilde{P}_{c_i}\nonumber\\
&+\alpha^2_iR_{i}(\nabla F_i(\eta_{i})-\nabla F_{i}(\zeta_{i}))\tilde{P}_{c_i}\nonumber\\
&-(\alpha_i+\alpha_i^2R_i)(\nabla F_i(\eta_{i})-\nabla F_{i}(\zeta_{i}))(\eta_{i}-\zeta_{i}))\nonumber\\
&+\sum\nolimits_{i\in\mathcal{N_L}}(-D_i\tilde{\omega}^2_i-\alpha_i(K_i-D_i)(\nabla U_i(\eta_{i})-\nabla U_{i}(\zeta_{i}))\tilde{\omega}_i\nonumber\\
&+\alpha^2_ia_i\tilde{d}_i^2-\alpha^2_ia_i(\eta_{i}-\zeta_{i})\tilde{d}_i-\alpha^2_i(\nabla U_i(\eta_{i})-\nabla U_{i}(\zeta_{i}))\tilde{d}_i\nonumber\\
&+(\alpha_i+\alpha^2_i)(\nabla U_i(\eta_{i})-\nabla U_{i}(\zeta_{i}))(\eta_{i}-\zeta_{i})).\label{ys}
\end{align}
Under Assumption 1, it follows from Lemma 2 in \cite{lip} that
\begin{equation}\label{af}
\begin{split}
\vert\nabla F_i(\eta_{i})-\nabla F_i(\zeta_{i})\vert
&\geq a_i\vert\eta_{i}-\zeta_{i}\vert,~\forall i\in\mathcal{N_G}\\
\vert\nabla U_i(\eta_{i})-\nabla U_i(\zeta_{i})\vert
&\geq -a_i\vert\eta_{i}-\zeta_{i}\vert,~\forall i\in\mathcal{N_L}.
\end{split}
\end{equation}
Moreover, according to Lemma 4 in \cite{lip}, the following facts hold under Assumption 2
\begin{equation}\label{bf}
\begin{split}
&b_i(\nabla F_i(\eta_{i})-\nabla F_i(\zeta_{i}))(\eta_{i}-\zeta_{i})\geq(\nabla F_i(\eta_{i})-\nabla F_i(\zeta_{i}))^2\\
&-b_i(\nabla U_i(\eta_{i})-\nabla U_i(\zeta_{i}))(\eta_{i}-\zeta_{i})\geq(\nabla U_i(\eta_{i})-\nabla U_i(\zeta_{i}))^2.
\end{split}
\end{equation}
Combing the facts \eqref{ys}-\eqref{bf} together gives the inequality \eqref{vde} presented below where 
\begin{equation}   
\begin{split}  
Q_i&=\left[              
\begin{array}{cccc}   
D_i&\frac{1}{2}\alpha_i(K_i-D_i)\\  
\frac{1}{2}\alpha_i(K_i-D_i)&\alpha_ib_i^{-1}+\alpha_i^2b_i^{-1}R_i-\alpha_i^2a_i^{-1}R_i
\end{array}
\right]\nonumber\\
&\hspace{7cm}\forall i\in\mathcal{N_G}\nonumber\\
Q_i&=\left[              
\begin{array}{cccc}   
D_i&\frac{1}{2}\alpha_i(K_i-D_i)\\  
\frac{1}{2}\alpha_i(K_i-D_i)&\alpha_ib_i^{-1}+\alpha_i^2b_i^{-1}+\alpha_i^2a_i^{-1}
\end{array}
\right]\nonumber\\
&\hspace{7cm}\forall i\in\mathcal{N_L}.\nonumber
\end{split}
\end{equation}
According to the Schur complement condition \cite{sch} and condition \eqref{sc}, matrix $Q_i$ is positive definite for any $i\in\mathcal{N}$. Therefore, $\dot{V}(\tilde{x})\leq 0$ from the equality \eqref{vd4}.

Define set $\Omega_2=\{\tilde x\in\Omega_1~|~V(\tilde x)\leq V(\tilde x(0))\}$ with $\tilde x(0)\in\Omega_1$ and $\tilde x(0)$ being bounded. We claim that $\Omega_2$ is compact and forward invariant with respect to system \eqref{cs}. In $\Omega_2$, the variables $\tilde\omega_\mathcal{G}$, $\tilde P_m$, $\tilde P_c$, $\tilde d$, $\tilde \phi$, $\tilde\gamma$, $\tilde z$, and $M\tilde\omega+\alpha\tilde\lambda$ are bounded  because of the non-negative quadratic terms in $V(\tilde x)$ and the boundedness of $V(\tilde x)$ in $\Omega_2$. $\tilde\xi$ is bounded by the definition of set $\Omega_1$ and Assumption 3. $\tilde\omega_\mathcal{L}$ is bounded in $\Omega_2$ because $\tilde\omega_\mathcal{L}=-D_\mathcal{L}^{-1}(\tilde d+C_{p_\mathcal{L}}T_p\text{sin}(\tilde\xi+\xi^*)-C_{p_\mathcal{L}}T_p\text{sin}(\xi^*))$ from system \eqref{cs} and $\tilde d$, $\text{sin}(\tilde\xi+\xi^*)$, $\text{sin}(\xi^*)$ are all bounded in $\Omega_2$. Hence, $\tilde\omega$ is bounded in $\Omega_2$. Then, the boundedness of $\tilde\lambda$ in $\Omega_2$ is obvious due to the boundedness of $M\tilde\omega+\alpha\tilde\lambda$ and $\tilde\omega$. Therefore, set $\Omega_2$ is bounded. 	In addition, let $\tilde x(t,\tilde{x}(0))=x(t,x(0))-x^*$, where $x(t,x(0))$ is the trajectory of system \eqref{cs} starting at $x(0)$. Since $\dot{V}(\tilde x)\leq 0$ in $\Omega_2$, $\tilde x(t,\tilde{x}(0))$ remains in $\Omega_2$ for $\forall t\geq0$. Hence, according to \cite{nl}, $\Omega_2$ is a compact forward invariance set in terms of system \eqref{cs}.

Now, consider the set $\Omega_3=\{\tilde x\in\Omega_2~|~\dot{V}(\tilde x)=0\}$. We claim that the largest invariance set of $\Omega_3$ only contains the point $\tilde{x}=0$ with respect to system \eqref{cs}. From \eqref{vd4}, we have $\tilde\omega=0$, $\tilde P_{m}=0$, $\tilde P_c=0$, $\tilde d=0$, and $\tilde\lambda=0$ in $\Omega_3$. Since $\tilde\omega=0$ and $\omega^*=0$, we have $\dot{\omega}=0$ and $\dot{\xi}=0$ in $\Omega_3$. According to \eqref{cs}, $\tilde P_{m}=0$, $\tilde P_c=0$, $\tilde d=0$, and $\dot{\omega}=0$ in $\Omega_3$ imply that
\setcounter{equation}{30}
\begin{align}\label{xi}
C_{p}T_p\text{sin}(\tilde\xi+\xi^*)&=C_{p}T_p\text{sin}(\xi^*).
\end{align}
Then, $\dot{\xi}=0$ and equation \eqref{xi} together indicate that $\xi=\xi^*$ in $\Omega_3$, because matrix $C_p$ has full column rank and $\vert \tilde\xi_{ij}+2\xi^*_{ij}\vert<\pi$, $\forall (i,j)\in\mathcal{L}_p$, in $\Omega_3$. Moreover, $\tilde\lambda=0$ indicates $\dot{\lambda}=0$ in $\Omega_3$. It follows that $L_c\tilde\phi=0$ in $\Omega_3$.  There accordingly exists a time-dependent scalar $\psi(t)\in\mathbb{R}$ such that $\tilde\phi=\psi(t)1_n$. Calculating $1_n^T\tilde\phi$ gives $1_n^T\tilde\phi=n\psi(t)$. By recalling \eqref{29a}, we can easily get $\psi(t)=0$. Thus, $\tilde\phi=0$, and $\dot{\phi}=0$ in $\Omega_3$. Combing $\dot{\phi}=0$, $\tilde{\lambda}=0$, and $\tilde{\omega}=0$ gives that $L_c\tilde\gamma=0$ in $\Omega_3$, which further implies that 
\begin{align}\label{gt}
\tilde{\gamma}=\tau(t)1_n^T
\end{align}
with time-dependent scalar $\tau(t)\in\mathbb{R}$. Left multiplying both sides of equation \eqref{gt} with $1_n^T$ gives that $1_n^T\tilde\gamma=n\tau(t)$. Moreover, from the fact specified in the proof of Theorem 1 that $1_n^T\gamma\equiv 1_n^T\gamma(0)$, we have $1_n^T\tilde{\gamma}\equiv 0$. Hence, $\tau(t)=0$, and $\tilde{\gamma}=0$ from \eqref{gt}. Finally, $\tilde\gamma=0$ implies $\dot{z}=0$, and hence $\tilde{z}=0$ in $\Omega_3$. Now, we can conclude that the largest invariance set of $\Omega_3$ in terms of system \eqref{cs} only contains the point $\tilde x=0$.

By the LaSalle invariance principle \cite{nl}, $\tilde x(t,\tilde{x}(0))$ with $\tilde{x}(0)\in\Omega_1$ approaches the largest invariance set of $\Omega_3$ as $t$ goes to infinity, i.e., $\lim_{t\to\infty}\tilde{x}(t,\tilde{x}(0))=0$. Therefore, $\lim_{t\to\infty}x(t,x(0))=x^*$, which implies that the equilibrium point $x^*$ of the closed system \eqref{cs} is asymptotically stable. $\hfill\blacksquare$

\textit{Remark 7:} Theorem \ref{the2} establishes a sufficient stability criterion for system \eqref{model} with the proposed control algorithm \eqref{ca}. It shows that the asymptotic stability of the closed-loop system \eqref{cs} relies on the selection of parameters $\alpha_i$ and $K_i$, $i\in\mathcal{N}$, that satisfy condition \eqref{sc}. In practice, the generator damping and load frequency sensitive coefficients $D_i$ are usually time-varying and hard to measure accurately \cite{ed}. However, like most results in distributed frequency control methods (e.g., \cite{both6,pd2,cl3,pd1,both1,both4,both2,both3}), directly checking this inequality requires the exact values of $D_i$ of all buses. To cope with this issue, we provide an alternative way to select $\alpha_i$ and $K_i$ in a distributed way in the case where $D_i$ is not exactly known. Although the exact values of $D_i$ are not available, they are bounded in practice, i.e., $D_i\in[D_i^{\text{min}},D_i^{\text{max}}]$. It is reasonable to assume that $D_i^{\text{min}}$ and $D_i^{\text{max}}$ are known and thus can be used to design control parameters. Then, $\varrho_{i}$ in \eqref{as} is upper bounded by
\begin{align}
\varrho_{i}^*
&=\frac{ b_iK_i^2}{4D^{\text{min}}_i}-\frac{b_iK_i}{2}+\frac{ b_iD^{\text{max}}_i}{4}+\frac{b_iR_i}{a_i}-R_i+\epsilon_i,~i\in\mathcal{N_G}\nonumber\\
\varrho_{i}^*&=\frac{ b_iK_i^2}{4D^{\text{min}}_i}-\frac{b_iK_i}{2}+\frac{ b_iD^{\text{max}}_i}{4}-\frac{b_i}{a_i}-1+\epsilon_i,~i\in\mathcal{N_L}
\nonumber
\end{align}
with arbitrary positive constants $\epsilon_i>0$, $\forall i\in\mathcal{N}$. In this case, if $\alpha_s^*$ satisfy
\begin{align}\label{al}
\alpha_s^{*}=\left(\max\limits_{i\in\mathcal{N}_{s}}\{\varrho_{i}^*\}\right)^{-1},~\forall s\in\mathcal{K}
\end{align}
condition \eqref{as} is guaranteed. Therefore, to determine parameters $\alpha_i,K_i$ satisfying \eqref{sc}, each bus $i\in\mathcal{N}$ can select its own $K_i$, $\epsilon_{i}$ at first, and meanwhile calculate $\varrho_{i}^*$ only based on its local information $a_i$, $b_i$, $\epsilon_i$, $K_i$, $R_i$, $D_i^{\text{min}}$, $D_i^{\text{max}}$. Then, for each control area, the maximum of $\varrho^*_{i}$ can be computed in~finite time by using the distributed max-consensus algorithm proposed in \cite{max} via communication between buses within the same control area. Finally, $\alpha^*_s$, $s\in\mathcal{K}$, can be determined following \eqref{al}, and $\alpha_i$ is selected accordingly based on \eqref{hi} for buses $i\in\mathcal{N}_s$ after a consensus is achieved. It should~be pointed out that, according to \eqref{sc}, the selection~of~$\alpha_i$~(or~equivalently $\alpha^*_s$) depends on the parameters of all buses within control area $s$. This means that $\alpha_i$ might need to be adjusted if a new bus is connected to area $s$. Hence, how to determine $\alpha_i$ by only using local parameters deserves attention.

\textit{Remark 8:} It follows from \eqref{ca} that the proposed distributed frequency control algorithm relies on the exact values of inertia $M_i$. In practice, the parameter $M_i$ is usually available with high accuracy, and thus can be used for the controller design \cite{ed}. In fact, as seen in Section \uppercase\expandafter{\romannumeral4} by simulation where we~use an estimated value of $M_i$ rather than its exact value for each generator in the proposed distributed controller, the controller \eqref{ca} appears to have robustness against parameter uncertainties of $M_i$, and asymptotic results are retained. However, how to theoretically guarantee the robustness of the distributed frequency controller with respect to uncertainties in the inertia values remains open and should be studied in the future.

\textit{Remark 9:} The designed distributed frequency control algorithm \eqref{ca} relies on information exchanges between the cyber-connected buses via a communication network. As argued in \cite{dos}, the power grids equipped with communication infrastructures may suffer from network attacks, e.g., denial-of-service (DoS) attacks and deception attacks, which may have negative impacts on the performance of the proposed algorithm. For example, if control gain $K_i$ is maliciously altered to violate the sufficient stability criterion \eqref{sc} under deception attacks, the asymptotic stability of the closed-loop system may not be guaranteed. Therefore, how to make the designed controller resilient against network attacks so as to guarantee the cyber-security of power systems is of great importance. This topic will be addressed in the future.

\textit{Remark 10:} For a single-area power system, the controller \eqref{ca} does not need $\gamma_i$ as well as $z_i$, and the dynamics of $\phi_i$ can be simplified into 
\begin{align}
\dot{\phi}_{i}=&-\sum\nolimits_{(i,j)\in\mathcal{L}_c}l_{c_{ij}}(M_i\omega_{i}-M_j\omega_j)\nonumber\\
&-\sum\nolimits_{(i,j)\in\mathcal{L}_c}l_{c_{ij}}(\alpha_i\lambda_{i}-\alpha_j\lambda_j),~i\in\mathcal{N}.\label{s}
\end{align}
In the vector form, \eqref{s} can be rewritten as 
\begin{align}
\dot{\phi}=L_c\alpha\lambda+L_cM\omega.
\end{align}
Since $L_c$ is the Laplacian matrix of the connected and undirected graph $G_{c}(\mathcal{N},\mathcal{L}_c)$, and $\omega^*=0$ at the steady state, $\lambda_i^*$ are identical for all buses $i\in\mathcal{N}$, where we note that $\alpha_{i}=\alpha_{j}$, $\forall i,j\in\mathcal{N}$. This means all buses reach the same incremental cost/utility values at the equilibrium point. Then, the optimal power allocation of all controllable units can be achieved in the system level as indicated in Remark 3.

\textit{Remark 11:} The assumption of positive load frequency sensitive coefficients, i.e., $D_i>0$, $\forall i\in\mathcal{N_L}$, in Section \uppercase\expandafter{\romannumeral2} is commonly used in the literature \cite{sp,bd}. Nevertheless, in practice, $D_i$ can be zero if the load at bus $i$ is frequency~independent. For this type of load buses, the corresponding controller \eqref{ca} can be obtained by setting $K_i=0$. The derived results in this paper will not be impacted. This is because the Lyapunov function defined for proving the closed-loop stability is independent of load frequency coefficients (see Theorem 2 for details), and thereby, applies to the cases which only require non-negative load frequency coefficients.

\section{Simulation Results}


\begin{figure}[t]
\centering
\epsfig{figure={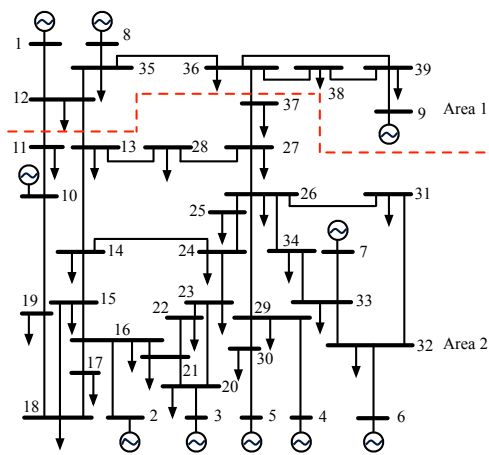},width=0.8\linewidth}
\caption{Diagram for the IEEE 39-bus system. Black solid lines: transmission lines, red dotted lines: boundary of two different control areas.}
\vspace{-0.3cm}
\label{figurelabel}
\end{figure}

\begin{table}[t]
\caption{Cost and Utility Function Coefficients}
\begin{center}
\setlength{\tabcolsep}{3.5pt}
\renewcommand\arraystretch{1.2}
\begin{tabular}{c|cccccccccccc}
\hline
\hline
Bus&1&2&3&4&5&6&7&8&9&10&Loads\\
\hline
$c_{1i}$&2.4&4&3.4&3&2.8&3.2&4&3.6&2.6&3&-3\\
\hline
$c_{2i}$&10.5&6.7&7.5&8.9&8.3&7.2&9.1&8&9
&6.5&9\\
\hline
$c_{3i}$&19.5&15&14.5&16.3&16.6&18.9&10
&17.9&11.1&13.8&12.5\\
\hline
\hline
\end{tabular}
\end{center}
\vspace{10pt}
\caption{Bus Parameters}
\begin{center}
\setlength{\tabcolsep}{3.3pt}
\renewcommand\arraystretch{1.2}
\begin{tabular}{c|cccccccccccc}
\hline
\hline
Bus&1&2&3&4&5&6&7&8&9&10&Loads\\
\hline
$M_i$&13&12.1&14.3&11.4&10.4&13.9&10.6
&9.7&13.8&16.8&$-$\\
\hline
$D_i$&1&0.8&1.1&1&0.9&1&1.2&0.8&0.9&1.1&1\\
\hline
$T_i$&0.3&0.4&0.35&0.3&0.33&0.37&0.4
&0.3&0.35&0.33&$-$\\
\hline
$R_i$&0.05&0.05&0.05&0.05&0.05&0.05&0.05
&0.05&0.05&0.05&$-$\\
\hline
\hline
\end{tabular}
\end{center}
\end{table}

\begin{figure}[t]
\centering
\includegraphics{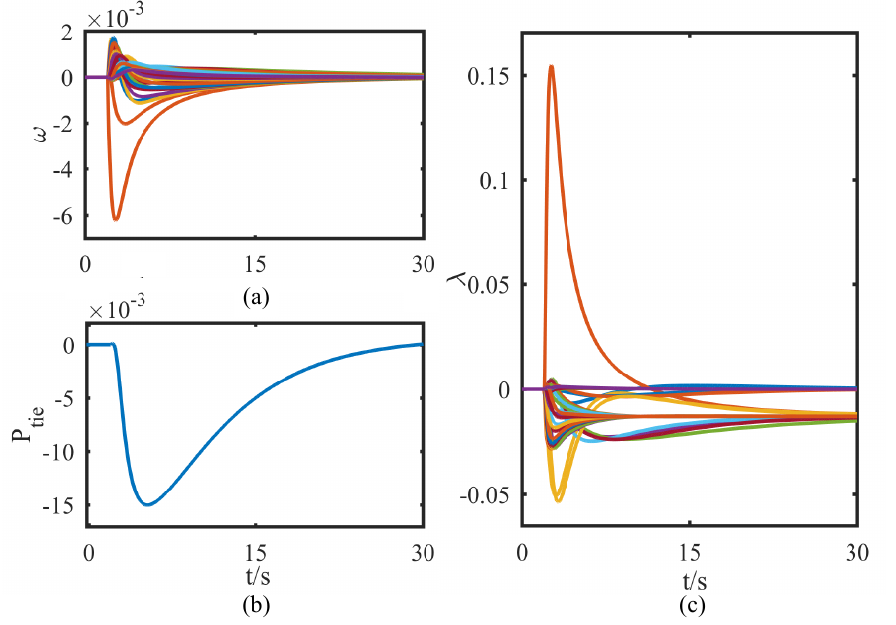}
\setlength{\abovecaptionskip}{-13pt}
\caption{State responses of the system.}
\vspace{0.2cm}
\epsfig{figure=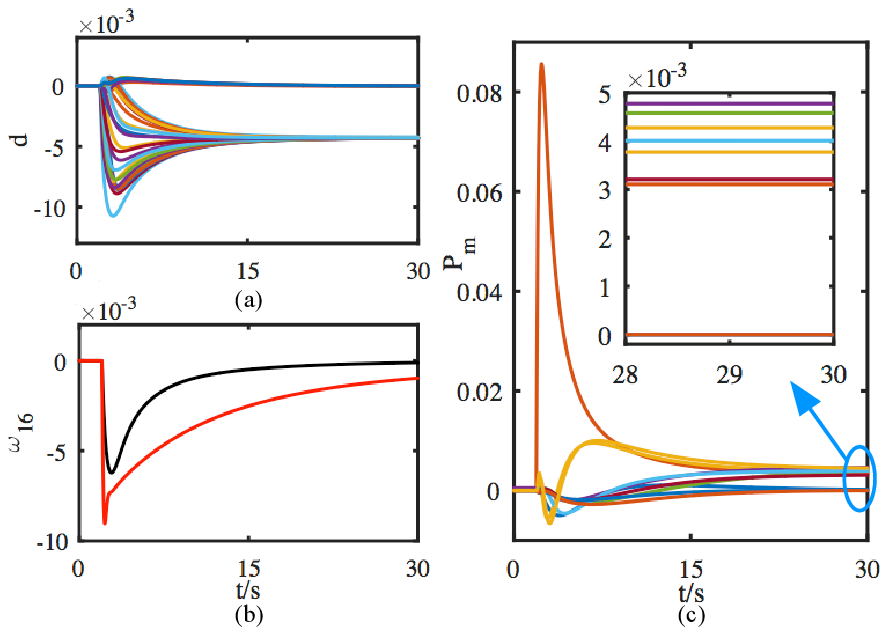,width=1\linewidth}
\setlength{\abovecaptionskip}{-7pt}
\caption{(a) Controllable load changes; (b) Frequency response at bus 16 (AGC: red line; the designed algorithm: black line); (c) Generator mechanical power changes.}
\label{figurelabel}
\vspace{-0.6cm}
\end{figure}

In this section, we use the IEEE 39-bus system to evaluate the effectiveness of the proposed control algorithm. We adopt parameters of the test system provided in \cite{prm}, where the system is divided into two control areas by the red dotted line as shown in Fig. 1. We assume that the power network initially operates at a nominal stable point, and adopt the quadratic generation cost function $F_i(P_{m_i})=\frac{c_{1i}}{2}P_{m_i}^2+c_{2i}P_{m_i}+c_{3i}$, $i\in\mathcal{N_G}$ (user utility function $U_i(d_{i})=\frac{c_{1i}}{2}d_{i}^2+c_{2i}d_{i}+c_{3i}$, $i\in\mathcal{N_L}$) for each generator (load) bus. The detailed cost/utility~function~coefficients are given in Table \uppercase\expandafter{\romannumeral1}. Particularly, the utility function coefficients are selected identically for all load buses for simplicity. The rest bus parameters in per unit on a base of 100 MVA are provided in Table \uppercase\expandafter{\romannumeral2}. Here, the generator damping and load frequency sensitivity coefficients are unknown but within range $[0.8,1.2]$. Hence, we set $\alpha_i=1$ and $K_i=1.2$, $\forall i\in\mathcal{N}$, such that condition \eqref{sc} is fulfilled. Further, we adopt the communication network with the same topology as the physical transmission network, and let $l_{c_{ij}}=-1$ for all communication links. To test the robustness of the proposed controller, we assume that the inertia of each generator is not exactly known. For simplicity, in the proposed distributed controller, we use the estimated values $M_i=12$ for each generator bus $i=1,2,\dots,10$, instead of using the exact values shown in Table \uppercase\expandafter{\romannumeral2}.


At time $t = 2$ s, a 0.13 p.u. (13 MW) load increase occurs at bus 16. The responses of bus frequencies, the deviation of the net tie-line power exchange between the two control areas from its scheduled value, and deviation of $\lambda_i$, $i\in\mathcal{N}$ from its original value are given in Fig. 2. It can be observed that the frequency and inter-area power exchanges are restored to their nominal values, which validates the effectiveness of the proposed control approach in frequency regulation. In addition, $\lambda_i$ converges to two values corresponding to the two different control areas. Then, according to Lemma \ref{lem1}, the total power mismatch between generation and demand is optimally shared among all generators and controllable loads. The controllable load changes and mechanical power changes of generators from their initial values are given in Fig. 3(a) and Fig. 3(c), respectively. It should be noted that, as the load increase only occurs within control area 2, the generators and controllable loads in control area 1 do not act in response to the disturbance at the steady state, i.e., their power changes and incremental cost/utility value changes all converge to zero. Further, due to the facts that the utility functions and incremental utilities are identical for all load buses in control area 2, the corresponding controllable load changes converge to the same values.

Fig. 3(b) compares the control performance between AGC and the proposed control method by showing the frequency at bus 16 under different control schemes.  AGC is implemented as in \cite{agc}, where the integral gain for the area control error (ACE) is chosen as 0.2 for the two control areas, and the~participation factors for generators in the same area~are~proportional to $c_{1i}$. Obviously, a smaller frequency nadir and faster convergence rate can be achieved by our method.

\section{Conclusion}

This paper has investigated the frequency regulation issue~of power systems by using a distributed frequency controller that can optimally coordinate active power outputs/consumptions of generators/controllable loads, and restore the nominal~frequency as well as the net tie-line power flows between control areas. Asymptotic stability of the closed-loop system under the proposed algorithm has been analysed with a nonlinear struc-ture preserving model, and a stability criterion of the system  under the proposed control algorithm on selecting parameters has been established. Furthermore, it has been shown that our controller is robust to parameter uncertainties. The simulation results have demonstrated the validity of our method.

\end{document}